\documentstyle[12pt,aasms4]{article}

\def\kmsec{\mbox{km~s$^{\rm -1}$}}

\def\BmV0{\mbox{(B-V)$^{\rm o}$}}
\def\VmK0{\mbox{(V-K)$^{\rm o}$}}
\def\MV0{\mbox{M$_{\rm V}^{\rm o}$}}

\def\Msun{\mbox{M$_{\odot}$}}

\lefthead{}
\righthead{}
\begin{document}

\title{ 
Neutron-Capture Elements in the Early Galaxy: Insights from 
a Large Sample of Metal-Poor Giants }

\author{
Debra L. Burris\altaffilmark{1},
Catherine A. Pilachowski\altaffilmark{2},
Taft E. Armandroff\altaffilmark{2},
Christopher Sneden\altaffilmark{3},
John J. Cowan\altaffilmark{4},
and Henry Roe\altaffilmark{5}}
\begin {center}
dburris@okc.cc.ok.us,
catyp@noao.edu,
armand@noao.edu,
chris@verdi.as.utexas.edu,
cowan@mail.nhn.ou.edu,
hroe@doty.berkeley.edu
\end {center}


\altaffiltext{1}{Oklahoma City Community College, Division of Science
and Mathematics, 7777 S. May Ave, Oklahoma City, OK 73159}
\altaffiltext{2}{NOAO, PO Box 26732, Tucson, AZ 85726-6732.
The National Optical Astronomy Observatories (NOAO) is operated
by the Association of Universities for Research in Astronomy, Inc.
(AURA) under cooperative agreement with the National Science Foundation.}
\altaffiltext{3}{Department of Astronomy and McDonald
 Observatory, University of Texas, Austin, TX 78712}
\altaffiltext{4}{Department of Physics and Astronomy, University of Oklahoma, Norman, OK 73019 }
\altaffiltext{5}{University of California at Berkeley, Berkeley, CA 94720}

\begin{abstract}
New abundances for neutron-capture ({\em n}-capture) elements in a large
sample of metal-poor giants from the Bond survey are presented.
The spectra were acquired with the KPNO 4-m echelle and coud\'e feed
spectrographs, and have been analyzed using LTE fine-analysis techniques
with both line analysis and spectral synthesis.
Abundances of eight {\em n}-capture elements (Sr, Y, Zr, Ba, La, Nd, Eu, Dy)
in 43 stars have been derived from blue
($\lambda\lambda$4070--4710~\AA, R$\sim$20,000, S/N ratio$\sim$100-200)
echelle spectra and red
($\lambda\lambda$6100--6180~\AA, R$\sim$22,000, S/N ratio$\sim$100-200)
coud\'e spectra, and the abundance of
Ba only has been derived from the red spectra for an additional 27 stars.

Overall, the abundances show clear evidence for a large star-to-star
dispersion in the heavy element-to-iron ratios.
This condition must have arisen from individual nucleosynthetic events
in rapidly evolving halo progenitors that injected newly manufactured
{\em n}-capture elements into an inhomogeneous early Galactic halo 
interstellar medium.
The new data also confirm that at metallicities [Fe/H]$\lesssim$--2.4,
the abundance pattern of the heavy (Z$\ge$56) {\em n}-capture elements
in most giants is well-matched to a scaled Solar System {\em r}-process
nucleosynthesis pattern.

The onset of the main {\em r}-process can be seen at [Fe/H]$\approx$--2.9;
this onset is consistent with the suggestion that low mass Type II supernovae
are responsible for the {\em r}-process.
Contributions from the {\em s}-process can first be seen in some stars with
metallicities as low as [Fe/H]$\sim$--2.75, and are present in most stars 
with metallicities [Fe/H]$>$--2.3.
The appearance of {\em s}-process contributions as metallicity increases 
presumably reflects the longer stellar evolutionary timescale of
the (low-mass) {\em s}-process nucleosynthesis sites.

The lighter {\em n}-capture elements (Sr-Y-Zr) are enhanced
relative to the heavier {\em r}-process element abundances.
Their production cannot be attributed solely to any combination of
the Solar System {\em r-} and main {\em s-}processes, but requires
a mixture of material from the {\em r}-process and from an additional
{\em n}-capture process which can operate at early Galactic time.
This additional process could be the
weak {\em s}-process in massive ($\sim$25~\Msun) stars, or perhaps
a second {\em r}-process site, i.e. different than the site that
produces the heavier (Z$\ge$56) {\em n}-capture elements.

\end{abstract}


\section{INTRODUCTION}

The oldest metal-poor halo stars are Galactic fossils that provide clues
to the conditions and populations of stars that existed early in the Galaxy's
history.  The chemical compositions of these halo stars are due to only a
few, perhaps one, earlier generations of stars.  Metal-poor stars provide
an opportunity to observe neutron-capture ({\em n}-capture) elements
(Z$>$30) produced in the unseen precursors to Population II, and through
their abundances, to deduce characteristics of the first 
Galactic stellar population.

The {\em n-}capture elements are produced through
both  slow ({\em s-}) and rapid ({\em r-}) {\em n}-capture processes.
The {\em s-} and {\em r-}processes are believed to occur at different sites.
The {\em r-}process requires a high neutron flux level (with many 
{\em n}-captures over a timescale of a fraction of a second)
thought to occur in supernova explosions,
while the {\em s-}process, which requires a lower neutron flux
(with a typical {\em n}-capture taking many years),
is generally thought to occur during the double-shell burning phase
of low (1-3~\Msun) and intermediate-mass (4-7~\Msun) asymptotic giant branch (AGB) stars.
In the Solar System, where the abundances of individual isotopes can be
determined (\cite{and82,and89,kap89,wis96}), the relative fractions of
elements produced by each process can be identified
(see Sneden et al. 1996, and Appendix A).

The production mechanisms for {\em n}-capture elements in metal-poor
stars have been the subject of some debate in the literature.
Many observational and theoretical studies 
(\cite{spi78,tru81,sne83,sne85,gil88},
Gratton \& Sneden 1991, 1994; \cite {sne94};
McWilliam et al. 19995a, 1995b; \cite{cow96,sne96}, Ryan et al. 1996)
have demonstrated
that the observed abundances of {\em n-}capture elements in metal-poor
stars are consistent with production via the {\em r}-process, and at least 
for the heaviest such elements, 
correspond to the scaled Solar System {\em r-}process signature
(\cite{gil88,cow95,sne96,sne98,cow97,cow99}).
Gilroy et al. was one of the first large surveys of these metal-poor
stars, and confirmed the operation of the {\em r-}process at low metallicity.
Gilroy et al. also revealed significant star-to-star scatter in the overall
abundance level of the {\em n-}capture elements with respect to iron for
stars with [Fe/H]$<$--2.0\footnote
{We adopt the usual spectroscopic notations that
[A/B]~$\equiv$~log$_{\rm 10}$(N$_{\rm A}$/N$_{\rm B}$)$_{\rm star}$~--
log$_{\rm 10}$(N$_{\rm A}$/N$_{\rm B}$)$_{\odot}$, and that
log~$\epsilon$(A)~$\equiv$~log$_{\rm 10}$(N$_{\rm A}$/N$_{\rm H}$)~+~12.0,
for elements A and B. Also, metallicity will be arbitrarily defined
here to be equivalent to the stellar [Fe/H] value.}.

The surveys of Beers, Preston, \& Shectman (1985, 1992) increased
many-fold the number of observed ultra-metal-poor stars (i.e. [Fe/H]$<$--3),
extending
the range of metallicity over which {\em n-}capture element abundance
patterns can be investigated.
Spectroscopy of a sample of 33 ultra-metal-poor stars by
McWilliam et al. (1995a, b) confirmed the large star-to-star scatter
in the [{\em n-}capture/Fe]
abundance ratios at low metallicity.
An extreme example is the {\em n-}capture-rich, ultra-metal-poor
([Fe/H]$\simeq$--3.1) star CS~22892-052 (McWilliam et al. 1995a, b;
Sneden et al. 1994).
Cowan et al. (1995) showed that this star's observed {\em n-}capture element
abundances display the detailed signature of the scaled Solar
System {\em r-}process abundances.  The case became even more
compelling when abundances were determined for additional {\em n-}capture
elements, including several elements (terbium, holmium,
thulium, and hafnium) which had never before been detected in metal-poor halo
stars (Sneden et al. 1996).

Other studies (c.f. Magain 1995, Fran\c{c}ois 1996, Mashonkina et al. 1999)
have found the case for an {\rm r}-process origin of the
{\em n}-capture elements to be less certain.  
Since the dispersion in {\em n}-capture element abundances compared
to iron is so large for stars with metallicities below [Fe/H]$<$--1.0,
much of the disagreement can be traced to the small number and particular
selection of stars included.
Even with the Gilroy et al. (1988) and McWilliam et al. (1995a, b)
surveys, the number of
metal-poor stars with well-determined {\em n-}capture abundances remains
small, particularly in the metallicity regime --2.5$<$[Fe/H]$<$--1.0.
(Edvardsson et al. 1993, Woolf et al. 1995, and Jehin et al. 1999
surveyed numerous stars with metallicities [Fe/H]$>$--1.0.)  The metallicity
range --3.0$<$[Fe/H]$<$--1.5 covers a critical transition during which
{\em s-}process elements begin to appear in the Galactic chemical
mix.  Fortunately, the Bond (1980; hereafter ``Bond giants'')
survey of metal-poor giants provides
a sample of relatively bright stars in the appropriate metallicity
range to investigate this transition, and to delineate the
history of enrichment of {\em n-}capture elements in the Galaxy.

In the present paper, we present {\em n}-capture element abundances
(specifically Sr, Y, Zr, Ba, La, Nd, Eu, Dy)
for 43 metal-poor Bond giants, and Ba abundances only for an additional
27 Bond giants on a uniform system of metallicity.  In \S 2 we
present observational data and in \S 3 the analysis of the abundances.
In \S 4 we discuss the observed abundances in the context of
nucleosynthesis and Galactic chemical enrichment, and in \S 5
we summarize our conclusions.

\section{OBSERVATIONS AND REDUCTIONS}

Observations of 43 Bond giants were obtained with the Kitt Peak National
Observatory 4-m Mayall telescope during observing runs in November, 1988,
and June, 1989.  The spectrograph was configured with the 31.6~l~mm$^{-1}$
echelle grating, a 226~l~mm$^{-1}$ cross dispersing grating blazed at
$\lambda$8500~\AA , and the UV fast camera (0.267 m focal length).
A CuSO$_{4}$
filter isolated the second order of the cross disperser.  The detectors,
800 x 800 pixel Texas Instruments CCDs designated ``TI2'' or ``TI3,'' were
binned two pixels in the spatial direction to reduce the effect of
readout noise.
The spectra were aligned nearly parallel to the columns of the CCD
to facilitate the extraction of one-dimensional spectra.  The stars
observed are sufficiently bright that sky subtraction was not needed.
The gratings were oriented to provide complete spectral coverage from
$\lambda$4070 to 4710~\AA\ in 18 overlapping orders.
Exposure times ranged from 1800s for the brighter stars to 7200s for
the fainter ones.
 
Calibration data include 20 bias and 20 flatfield frames taken each night,
as well as an overscan region on each CCD frame.  Data reductions followed
standard IRAF\footnote{
IRAF is distributed by the National Optical Astronomy Observatories,
which are operated by the Association of Universities for Research
in Astronomy, Inc., under cooperative agreement with the National
Science Foundation.}
 procedures.
Each set of calibration frames was combined to remove cosmic ray events and to
reduce the noise.  For each data image, the bias level determined from
the overscan region and the corrected, average bias frame were subtracted.
The image was then divided by the bias-corrected, combined, normalized,
flatfield frame.
The IRAF task {\it apscatter} was used to subtract scattered
light in each echelle
image, and the one-dimensional spectra were extracted.  
A list of stellar lines whose wavelengths 
are well determined in the solar spectrum was used to determine the wavelength
calibration, and the continua were normalized to unity.
Spectra of a Th-Ar comparison lamp were also obtained at the beginning and
end of each night.
The resolving power of the spectra, measured from the FWHM (0.23~\AA )
of the Th-Ar lines near 4600 \AA\, is R$\sim$20,000.  The S/N ratio
per pixel varies from a few hundred (typically 200)
in the orders at the red end of the spectrum to about 100 in the orders
at the blue end.
 
The same 43 giants, as well as 27 additional Bond giants, were also observed
using the coud\'{e} feed telescope
and spectrograph to obtain spectra of the red $\lambda$6141 Ba II
feature.  Because the stars are red and the gratings and detectors
have greater efficiency in the red, we were able to use the smaller
(0.9-m) telescope for these observations.  Spectra were obtained in
November, 1988, and in May, 1989.  The spectrograph was
configured with the large collimator, the 632~l~mm$^{-1}$ ``A'' grating,
camera 5 (1.08~m focal length), and the ``TI3'' detector.
Exposure times ranged from a few minutes for the brightest stars to
2 hours for fainter stars.
Calibration procedures were similar to those described for the 4-m echelle
observations.  The resolving power is slightly higher, R$\sim$22,000, from
measured Th-Ar line FWHM of 0.28 \AA .
The S/N ratio is typically between 100 and 200 per pixel.

\section{ABUNDANCE DETERMINATIONS}

The abundance analyses were performed with the LTE spectral line
analysis code MOOG (Sneden 1973), adopting the model atmospheres
of Pilachowski et al. (1996, hereafter PSK) for most stars.
Those models were computed with the atmosphere code of Gustafsson
et al. (1975). PSK adopted effective temperatures (T$_{eff}$)
from calibrations of $V-R$ and Str\"omgren $b-y$ photometric indices.
The gravity (log~g) was computed from the mean of up to three independent
estimates: from the star's absolute magnitude (as deduced from its
Str\"omgren c$_{\rm 1}$ index); from its implied position in the M92
color-magnitude diagram; and from an average T$_{eff}$ vs. log~g
relationship for metal-poor giants gleaned from other studies in the
literature.  The microturbulent velocity (v$_{t}$) was estimated
from a correlation of T$_{eff}$ and (v$_{t}$) values for metal-poor
giants, except that for horizontal branch stars, a uniform
value of v$_{t}$~= 1.8~\kmsec\ was adopted.  In a few cases, PSK made
small adjustments to the T$_{eff}$, log~g, and v$_{t}$ parameters
through abundance computations from their line data.  See PSK
for a more complete description of the derivation of model atmosphere
parameters for our stars.
For stars not included by PSK, models were adopted as
noted in Table 1.  A new model atmosphere was generated for HD~29574
in an attempt to improve the fit of the Ti and Fe spectral features
from the models of PSK (see below).

The final model atmospheric parameters adopted for each star are
listed in Table~1, where [M/H] is the metallicity with respect to
the Sun of the models used in the analysis.
Abundance uncertainties due to errors in the model
atmospheres and atmospheric parameters were discussed in more detail
in Sneden et al. (1996).

The atomic transitions used in the analysis and the atomic line data
were adopted primarily from the extensive work of Sneden et al. (1996). 
Our choice of {\it n}-capture elements was somewhat restricted, being
limited by the observed wavelength ranges (4070$<$$\lambda$$<$4710~\AA ,
and 6100$<$$\lambda$$<$6141~\AA ), and by the intrinsic weakness of most
{\it n}-capture element transitions in low-metallicity stars.
No spectral features in the range from $\lambda\lambda$4225-4280~\AA\
were included due to the contamination from CH G-band features
in some stars.
Furthermore, since our chief goal was to trace the evolution of {\it n}-capture
elements as a function of stellar metallicity, we included only
elements that appear in our spectra over a wide range of [Fe/H]. 
Thus we did not include some elements (e.g. Sm and Ce, which do appear
in more metal-rich stars) because they become undetectable in lower metallicity
stars at our resolution and S/N ratio.
The lines included in our analysis and references to the atomic data are
listed in Table~2.

Before deriving abundances of the {\em n}-capture elements, 
we determined abundances of Ti and Fe from up to 10 transitions
of singly ionized species in the $\lambda\lambda$4450--4700~\AA\ region
of our blue echelle spectra. Our purpose was to establish whether the PSK
model atmospheres employed with our echelle spectra here yield metallicities 
in reasonable agreement with the PSK results, with particular attention to
the adopted microturbulence values, which PSK could not
determine from their spectra.  In Figure~1 we correlate 
abundances derived from the newly-measured Fe~II and Ti~II lines
with [Fe/H]$_{\rm PSK}$ and [Ca~I/H]$_{\rm PSK}$.
The iron abundances compared in the top panel are in excellent
accord: $<\delta$[Fe/H]$>$~=~+0.06~$\pm$~0.02 ($\sigma$~=~0.16; 43 stars),
in the sense Fe~II~{\em minus}~Fe$_{\rm PSK}$, with no trend
with metallicity.
Overall, these abundance correlations confirm that the PSK models and
atmospheric parameters are suitable for use with the echelle spectra
considered here, and provide reliable abundances.  With the exceptions
noted below, the microturbulence values adopted from PSK were confirmed
by our Fe~II and Ti~II equivalent widths.

Our [Ti/Fe] overabundances are, however, 
somewhat larger than are those of [Ca/Fe] from PSK:
$<$([Ti II/H]--[Fe/H]$_{\rm PSK96}$)$>$~= +0.45~$\pm$~0.03 ($\sigma$~=~0.20),
as compared with $<$([Ca~I/H]--[Fe/H]$_{\rm PSK96}$)$>$~= +0.23~$\pm$~0.02,
($\sigma$~=~0.12). The Ti II lines in many of our more metal-rich and 
cooler stars are quite strong, much more so than are the Fe II lines.  
For most of our stars the mean Ti II equivalent widths were $>$100~m\AA, 
and the strongest of these lines often exceeded 180~m\AA.
Thus they lie on the flat part of the curve-of-growth, and abundances
derived from them are sensitive to choices of v$_{t}$.  Such very strong
lines might not be reproduced very well by our model 
atmospheres and adopted analysis procedures.
For the nine program stars with the weakest Ti II equivalent widths 
($\lesssim$130~m\AA), $<$([Ti II/H]--[Fe/H]$_{\rm PSK96}$)$>$=+0.26,
in much better agreement with the $<$([Ca I/H]--[Fe/H]$_{\rm PSK96}$)$>$
from the whole sample.

The majority of the {\em n}-capture transitions which we used are blended with
other features; therefore, we chose to perform spectral syntheses rather
than to determine the abundances from single-line, equivalent width analysis.   
Figure~2 shows examples of the synthetic spectra for HD~63791.
The lines were fit by synthesizing spectra with a range of element/iron ratios,
typically ranging from one-tenth solar to three times
solar, then adjusting the element/iron ratios to bracket the best fit.
With two exceptions, we were able to derive the {\em n}-capture element
abundances from at least two detected lines. In HD~232078 and
BD~+30~2611, spectra of the $\lambda$4077 \AA\ region were not
usable due to low S/N ratio and line blending.
In these two stars, the Sr and Dy abundances are from single lines.  
For the unblended lines Nd~II~$\lambda$4446~\AA\ and $\lambda$4462~\AA\
we were able to employ equivalent width analysis.  These lines were measured 
with the equivalent width subroutine of the SPECTRE code
(Fitzpatrick \& Sneden 1987\nocite{FS87}), and the abundances were
determined directly by MOOG.  These abundances were then averaged
with the synthesis abundances obtained for the Nd~II~$\lambda$4358~\AA\
and $\lambda$4109~\AA\ lines to obtain a final Nd abundance.

The elements Ba, La, and Eu must be given special attention, since they
are subject to isotopic and/or hyperfine splitting which can affect the
final abundances.  These effects were treated in the same fashion 
as in Sneden et al. (1996).
Fortunately, abundances derived from the Ba~II $\lambda$6141 line
are not as sensitive to the distribution of Ba isotopes, which differ
depending on whether the barium is produced by the {\em r}-process
or the {\em s}-process.  But even for the $\lambda$4554 line, the 
assumption of a different isotopic distribution can change the
derived abundance by up to only $\sim$0.1~dex when the feature is strong.

The Ba abundance is also sensitive to the microturbulent
velocity because it is derived from strong lines, with equivalent widths
$\ge$60~m\AA\  (most other lines used in this analysis are typically weak).
Several of the stars showed a difference between the Ba abundance derived
from the $\lambda$4554~\AA\ and the $\lambda$6141~\AA\ features,
in the sense that the abundance derived from the $\lambda$4554~\AA\
line was a few tenths of a dex larger than that from the
$\lambda$6141~\AA\ line.
The $\lambda$4554~\AA\ line is much stronger and more sensitive to
microturbulence than the $\lambda$6141~\AA\ line.
Microturbulence was not well determined in the 
PSK study because that study made use of only weak lines.
The Ba abundance disparities were eliminated by
raising the microturbulence slightly ($\le$0.2~km~s$^{-1}$) from
the PSK value in the few stars where the two lines disagreed.

The Sr II lines used in our analysis are also sensitive to microturbulence.
Increasing the microturbulence to bring the two Ba II lines into agreement
also decreased the Sr abundance in these same stars by $\simeq$~0.15~dex.
Because the Sr II lines become so strong in stars cooler than 4500K,
we were unable to derive reliable Sr abundances in such stars.  Sr
abundances in warmer stars appear to be more reliable.

Tables~3a and b present our final {\em n-}capture element abundances.
Table~3a gives the metal-to-iron ratios [m/Fe] and Table~3b
gives the absolute log~$\epsilon$ abundances.
The total sample of stars included in Tables~1 and 3 is 70; Eu equivalent
widths from Gilroy et al. (1988) were used to redetermine abundances
for eight stars (noted in Table 1) for which new blue spectra were not
available.

Many of the Bond giants have been studied by previous authors,
allowing us to compare our results with the literature.
For the six stars in common with Gilroy et al. (1988), our analysis
results are in agreement, with a scatter of about 0.20 dex.
Our study also includes four stars in common with Gratton \& Sneden (1994), and
for most elements our log~$\epsilon$ values are higher by about 0.17~dex.

For the five stars in common with the study by McWilliam et al. (1995b)
and McWilliam (1998), we find a mean difference in [Ba/H] of +0.21 dex,
in the sense this paper {\it minus} McWilliam.  This difference does not
result from the measured line strengths, since the equivalent widths show
generally good agreement.
Hyperfine/isotopic line structure assumptions cannot be the cause
of the discrepancy, as they are the same in here and in McWilliam (1998).
Examination of the
McWilliam et al. (1995b) model atmosphere parameters compared to ours
for the stars in common suggest that the primary sources of the discrepancy
in the Ba abundances are differences in the adopted microturbulences.
Our microturbulence values are, on average,
lower by 0.75~km~s$^{-1}$.  Adopting the McWilliam et al. microturbulences
with our models brings the abundances into good agreement.
Since our spectra are typically
of higher signal-to-noise ratio than those available to McWilliam et al.,
and our microturbulence values are supported by consistency of the
Fe II, Ti II, and Ba II abundances from different lines, as described
above, we retain our original microturbulence values.
For other elements (which are less sensitive to microturbulence),
we agree on average to within $\pm$0.2 dex, although the number of stars
in common for these other elements is often only two or three.
There appears to be no systematic difference between our abundances and
those of McWilliam et al.  Overall, our abundances appear to be in
good agreement with other published values.

\section{{\em N}-CAPTURE ELEMENT ABUNDANCES}

\subsection{Star-to-Star Variations}

We first consider whether the star-to-star scatter in the
[{\em n}-capture element/Fe] abundance ratios seen by
Gilroy et al. (1988) and by McWilliam (1998) from poorer quality
spectra is also present in our data.  
Figure~3 compares the spectra of BD~+9~2870 and HD~6268, two stars
with similar atmospheric parameters and metallicity
($<$T$_{eff}$/log{\it g}/v$_{t}$/[Fe/H]$>$=4650/1.50/1.65/--2.4).
The spectral region displayed in the top panel shows that the Fe~II,
Cr~II, and Ti~II lines have nearly identical strengths in the two stars.
The second and third panels display regions containing {\em n}-capture lines.
Despite the close match of the Fe-peak lines,
ionized transitions of the {\em n}-capture elements Eu, La, Dy, and Nd
are 4-10 times stronger in HD~6268 than in BD~+9~2870.
All of the {\em n}-capture elements shown here vary
together: the strengths of features of La, Eu,
Dy and Nd are always greater in HD~6268 than in BD~+9~2870.
The {\em n}-capture element abundances must be
different in these two stars.

Star-to-star differences in the {\em n}-capture element abundances
are illustrated in Figure~4, in which we plot 
the ratios [Eu/Fe], [Dy/Fe] and [Dy/Eu] versus metallicity for 
the Bond giants.  Eu and Dy are both produced
primarily by the r-process in the Solar System (see Appendix~A
and Table~4).  For HD~6268 and BD~+9~2870
(these stars, whose spectra are shown in Figure~3,
are plotted with special symbols in Figure~4) the derived
Eu and Dy abundances are, as expected, much greater in the star
with stronger {\em n}-capture lines.
The ratio of [Dy/Eu] (bottom panel), however, is the same in these two stars.
The abundances of Dy and Eu vary together in all of the stars included
in our analysis, and the [Dy/Eu] ratio does not correlate with the
iron abundance. 
The individual [Eu/Fe] and [Dy/Fe] abundances vary by up to an
order of magnitude from solar, but this large scatter is not
seen in [Dy/Eu], as it would if caused by observational error.
The [Dy/Eu] ratios cluster tightly around the solar
ratio for the whole metallicity range studied.
These results, particularly the correlation between the Dy and Eu
abundances, indicate that the scatter of the {\em n}-capture elements'
abundances with respect to iron is real in metal-poor halo
stars and is not the result of observational error.

The behavior of all the heavy elements in our study
with metallicity is shown in Figure~5. 
In addition to our data, indicated by filled circles, we have included
data from McWilliam et al. (1995a, 1995b) and McWilliam (1998),
Jehin et al. (1999), and Edvardsson et al. (1993) and Woolf et al. (1995).
These particular studies were selected because they each include
at least 20 stars, they include hyperfine splitting in the calculation
of synthetic spectra for Ba and Eu, and together they cover the 
range of metallicity from --3.5$\le$[Fe/H]$\le$+0.5.
These studies were selected to provide a sample of stars covering this wide
metallicity range while minimizing observational scatter and systematic error.
The McWilliam et al (1995a,b) sample contains two known or suspected CH
stars with {\em s}-process enrichments, which we have eliminated from
further consideration.
These stars show characteristically high abundances of traditional
{\em s}-process elements, particularly Ba,
compared to other stars of similar metallicity.
We have retained the {\em r}-process-rich star CS~22892-052.

Our abundances confirm the trend of increasing scatter in the 
[{\em n}-capture/Fe] element ratios with decreasing metallicity noted
previously by Gilroy et al. (1988) and McWilliam (1998).
As metallicity increases, the scatter in the [{\em n}-capture element/Fe]
ratios declines dramatically.  At low metallicity, the scatter is
astrophysical in origin and not the result of observational uncertainty.  
At higher metallicities the scatter essentially disappears.
We agree with earlier studies that such
star-to-star variation is most easily explained as resulting from 
local nucleosynthetic events.  Interstellar matter undergoing star formation
closer to a site for {\em r}-process nucleosynthesis, presumably a 
supernova, might be enriched while other interstellar matter might not. 
We stress again that the observed star-to-star scatter appears in the total 
abundance level with respect to iron and not in the relative {\em n}-capture
abundances. The relative
abundances are similar in all the extremely metal-poor halo stars,
and similar to the Solar System {\em r}-process-only proportions (see Table~4).

\subsection{Abundance Trends with Metallicity}

In this section we examine more closely the abundances of individual
{\em n}-capture elements as a function of metallicity.
We look first at the behavior of [Ba/Fe], since
Ba abundances are available for stars of
lower metallicity than most other {\em n}-capture elements because
of its relatively higher abundance and its strong lines.
The fall in the [Ba/Fe] ratio at very low metallicity
([Fe/H]$<$-2.5), which has been noted in other studies and is
also seen in our data, is thought to result from the changing nature
of the nucleosynthesis sources for Ba. In the Solar System,
Ba is overwhelmingly produced in the main {\em s}-process
(see Table~4 and Appendix A) in low mass (1-3~\Msun) stars
with a small ($\sim$15\%) component from the {\em r}-process.
The initial production of Ba at the lowest metallicities, however, is
most likely by the {\em r}-process alone (Truran 1981), occurring in
supernovae resulting from more massive and hence more
rapidly evolving stars.

The [Ba/Fe] panel of Figure 5 shows that the transition between
a pure {\em r}-process production of Ba and production dominated
by the main {\em s}-process occurs gradually in the metallicity
regime from --2.75$\le$[Fe/H]$\le$--2.2.
The Ba abundance already approaches a solar [Ba/Fe] ratio in
some stars with metallicities as low as [Fe/H]=--2.75, hinting at
the presence of some 
{\em s}-process material in some stars even at this low metallicity.
A metallicity of [Fe/H]=--2.75 for the first appearance of material
of {\em s}-process origin is significantly lower than previously indicated. 
Based on the [Ba/Fe] data alone, however, we cannot discriminate between
the appearance of {\em s}-process material and a scatter in the
[{\em r}-process/Fe] ratio.

For most of the {\em n}-capture elements, the [{\em n}-capture/Fe]
ratios rise above the solar value (dotted line in Figure~5) before
falling back to a solar ratio at a metallicity near [Fe/H]=--1.0.
These high [{\em n}-capture/Fe] ratios are a reflection of the time
delay for the production of the bulk of the iron in the Galaxy.
While high mass Type~II supernovae are responsible for some of the
iron nucleosynthesis (particularly at very low metallicities),
Type~Ia supernovae (with longer evolutionary timescales due to lower
mass progenitors) are presumably responsible for the majority
of Galactic iron production (e.g. Matteucci \& Greggio 1986). 
The observed [{\em n}-capture/Fe] ratios for more metal-rich stars
(i.e. [Fe/H]$\simeq$--1.0) tend downward due to the
continued increase in Galactic iron production.
We caution that disk stars may behave in a different manner than
halo stars, and comparisons between these populations could be dangerous.
However, the downward trend as a function of increasing metallicity is
evident for all of the {\em n}-capture elements observed except Y,
including the r-process elements Eu and Dy and the
s-process element Ba, in all of the halo and disk stars where data are
available.  The disk and halo populations overlap in metallicity
near [Fe/H]$\simeq$--1, and the mix of populations does not appear to cause an
increase in the scatter of the [{\em n}-capture element/Fe] ratios.
The onset of the bulk iron production must have occurred
at a higher metallicity and presumably at a later time
than the onset of Galactic main {\em s}-process production.
 
Sr is also produced primarily by the {\em s}-process in the Solar System.
Like [Ba/Fe], [Sr/Fe] is low in very metal poor stars, rising
above a metallicity [Fe/H]=--3.0 to extend above the solar [Sr/Fe] ratio,
and then turning over in the abundance ratio with the increasing
formation of Galactic iron.
The scatter also decreases at higher metallicities as it does for [Ba/Fe].
However, the scatter appears to be larger in the [Sr/Fe] data, particularly at 
very low metallicity, than in the [Ba/Fe] data, as noted previously
by Ryan et al. (1996).
This might reflect the influence of an additional or different
production mechanism for Sr at very low metallicity (see discussion in
\S 4.4) or perhaps somewhat less certain observational
data for this element at the lowest metallicities. 

The abundance trends for the other elements shown in Figure~5 are
in general the same as shown for Ba and Sr, but with much less
complete data sets. Thus, for example, [Y/Fe] and [Zr/Fe] do not
show the rapid increase in abundance at low metallicities, but there are,
in fact, not much data for those elements at the lowest metallicities. 
We have already seen that Eu and Dy are correlated 
at the lowest metallicities, although there is much less data available
for Dy at higher metallicities.  (Dy was not included by Edvardsson
et al. 1993 or Woolf et al. 1995, or by Jehin et al. 1999.)
The turnover in [m/Fe] at increasing iron abundance and the increased
scatter at lower metallicities is exhibited by all of the elements
shown in Figure~5 (except Y).
Although the abundance data for Eu and Dy are limited or non-existent
at [Fe/H]$\simeq$--3.0, these elements do not seem to share the characteristic
of a decline in abundance relative to iron at [Fe/H]$<$--2.4 seen
in the traditional {\em s}-process elements Ba and Sr.
Indeed, as we will see later, the average abundance of these elements
at [Fe/H]=--2.8 is still well above solar, even when the {\em r}-process
rich star CS~22892-052 is excluded from the average.

\subsection{Heavier N-capture elements Ba--Dy }

The {\em r}-process element Eu has frequently been contrasted with Ba,
made predominantly in the {\em s}-process in solar material
(see Table~4 in Appendix~A and Figure~11).
The ratio of these two elements is often used to identify
{\em n}-capture nucleosynthesis mechanisms.
The logarithmic abundances of the two elements are plotted in Figure~6,
following Woolf et al. (1995).
In addition to our giants, we have included the same literature data sets
as in Figure~5.
Superimposed on the data are a dotted line for the scaled Solar System
{\em r}-process-only abundance ratio and a solid line for the scaled
total Solar System abundance ratio.  
For the lowest metallicities (i.e. log~$\epsilon$(Ba)$<$0.0),
the [Ba/Eu] ratio is constant and equal to the scaled 
Solar System {\em r}-process-only abundance ratio.  

Eu is made only by the {\em r}-process and Figure~6 clearly indicates
that at low metallicities (certainly at or below [Fe/H]=--2.75),
Ba is made in the same way.
This suggestion has been made previously (Spite \& Spite 1978; Truran
1981) based upon data for a few individual, metal-poor stars.
Sneden et al. (1996, 1998) and Cowan et al.  (1999) showed that
the well-studied ultra-metal-poor star CS~22892--052
([Fe/H] = --3.1), for example, has a {\em n}-capture
signature that is totally consistent with 
{\em r}-process-only {\em n}-capture nucleosynthesis.
Here we observe the same (solar {\em r}-process-only) ratio of Ba/Eu 
in many stars at low metallicity, suggesting that Ba and Eu are 
systematically made
together in the {\em r}-process during the early history of the Galaxy.
Other suggestions (e.g., an early Galactic {\em s}-process origin) 
for the Ba/Eu ratio in these stars seem unlikely. (See also McWilliam 1998.) 
The site(s) which produced Ba and Eu in the Solar System {\em r}-process-only
ratio in these very metal-poor halo must have evolved rapidly during the
early history of the Galaxy to synthesize the {\em n}-capture elements
while the Fe abundance was still very low (see also Raiteri et al. 1999).

Referring again to Figure~6, we note that the Ba/Eu abundance ratio changes
near log~$\epsilon$(Ba)=0 from the characteristic {\em r}-process value
to the value found for the total Solar System mixture of {\em r}-process
plus {\em s}-process contributions.  As the data in Table~3a indicate, this 
corresponds to a stellar metallicity range from --2.4$<$[Fe/H]$<$--2.1. 
At this metallicity, the ratio of Ba/Eu increases due to 
increased production of Ba, but not Eu. For stars with [Fe/H] above
this metallicity, the [Ba/Eu] ratio seems to reflect substantial
production of Ba by the {\em s}-process.  This transition marks
the main onset of Galactic {\em s}-processing at this
metallicity, sometime after the earlier onset of {\em r}-process synthesis.
This delay in {\em s}-processing is consistent with a longer stellar 
evolutionary timescale typical of low-mass (1--3 M$_{\odot}$) stars, thought to 
be the site for this type of {\em n}-capture synthesis (Gallino et al. 1998). 
The gradual rise in the Ba/Eu ratio suggests that the s-processing
sites must include a range of stellar masses.
Some {\em s}-process material may already be present at [Fe/H]=--2.75 in a
few stars, but the bulk of the {\em s}-processing is delayed until
near [Fe/H]=--2.2.  
The change in the Ba/Eu ratio near [Fe/H]=--2.2 is probably caused by
the (main) {\em s}-process occurring in the most massive 
(perhaps 2-3 M$_{\odot}$) of these low-mass stars, with additional
{\em s}-process contributions coming from increasingly lower
mass stars as the Galaxy ages.
The gradual rise in the [Ba/Eu] ratio over several tenths of a dex
in [Fe/H] is inconsistent with s-process production being limited
only to stars of $\sim$1~\Msun , which take 10$^{10}$ years to evolve
to the AGB stage, and may indicate local mixing effects and a range
of stellar mass progenitors for the main {\em s}-process production.
Somewhat more massive AGB progenitors (2--3 \Msun) may deposit
their ejecta earlier than the more
uniformly distributed and more numerous low-mass (1--2 \Msun) AGB stars.

The Ba and Eu abundances in Table~3b can also be used to estimate the
fraction of Ba produced by the {\em r}- and {\em s}-process in each star,
using the assumptions that Eu is produced only by the {\em r}-process
and that the {\em r}-process ratio of Ba/Eu is given by the {\em r}-process-only
fractional
abundances in the Sun.  Between 50--80\% of the Ba in most stars in our
halo sample with [Fe/H]$\ge$--2.0 is produced by the {\em s}-process.
Detailed analyses of the abundances of the metal-poor star HD~126238,
with [Fe/H]=--1.7 (Cowan et al. 1996), found that only a small percentage
(20\%) of the Ba could be attributed to the {\em s}-process in that star.
Our sample includes three other stars with metallicities near [Fe/H]=--2.0
which have relatively low {\em s}-process fractions (i.e. $<$40\%).
Cowan et al. (1996) suggested that at a metallicity of [Fe/H]=--1.7,
the {\em s}-process fraction was coming only from the most massive AGB sites.
Only at higher metallicities ([Fe/H]$>$--1.5) do most of the stars
have Ba {\em s}-process fractions at or near the solar value of 80\%.
The relative paucity of AGB stars with masses near $\simeq$3\Msun\
might well account for some of the dispersion in the [Ba/Fe] ratio
in stars at intermediate metallicity (--2.0$<$[Fe/H]$<$--1.0).
More metal-rich stars, formed after a time when more numerous and
more ubiquitous lower mass ($\simeq$1.5-3 \Msun) AGB stars are contributing
{\em s}-process elements, are increasingly uniform in their [Ba/Fe] ratio.
The major (low-mass, perhaps 1-1.5~M$_{\odot}$) sites
have not evolved to produce {\em s}-process elements until such time
as the Galactic iron abundance has reached [Fe/H]$\sim$--1.0.
At this metallicity, the Galaxy contains a mixture of different
stellar populations which may well confuse the interpretation
of Galactic chemical enrichment history.

The appearance of major phases of production of heavy elements in the
Galaxy can be seen by considering the abundances of {\em r}- and 
{\em s}-process element fractions separately.  In Figure~7, we plot
the {\rm r}-process-only and {\em s}-process only abundances of
barium as a function of [Fe/H], following Raiteri et al. (1999).
Note that the ``solar'' ratios in Figure~7 are not the total [Ba/Fe]
ratio, but the portions of Ba attributed to just the {\em r}- or
{\em s}-process.  The {\em r}-process-only
[Ba/Fe] ratio is just equal to the [Eu/Fe] ratio, since Eu is
produced almost entirely (97\%) by the {\em r}-process, and the Ba/Eu
{\em r}-process ratio is fixed at the {\em r}-process-only ratio.
For stars more
metal-poor than [Fe/H]$<$--2.75, for which there is no main {\em s}-process
contribution, the total Ba abundance can be used directly.  The
{\em s}-process-only barium abundance is simply what is left after
the {\em r}-process Ba is subtracted from the total observed abundance.
In those cases where the computed number of {\em r}-process-only barium atoms
exceeds the total number of barium atoms observed,
we have set the {\em r}-process-only
barium abundance to the total observed value and the {\em s}-process-only
value to zero (i.e. log~$\epsilon$(Ba$_{s-process}$)$\ll$0).

In the upper panel of Figure~7, {\em s}-process Ba first appears
at a metallicity [Fe/H]=--2.75, with the bulk of {\em s}-process production
delayed until [Fe/H]$\sim$--2.4,  consistent with the shift in the Ba/Eu
ratio seen in Figure~6.  The downwardly directed arrows in the upper panel
of Figure~7 mark the metallicities of stars which do not appear to contain
{\em s}-process material.
The existence of a wide (in metallicity) transition
region during which the Galactic {\em s}-processing gradually appears
may also explain why there might have been a ``controversy'' in the past,
with some authors suggesting on the basis of a single star that there was
significant {\em s}-, rather than only {\em r}-process production early 
(i.e., at low metallicity) in the Galaxy.

While the heavy elements (Z$\ge$56) at low metallicity are produced only
by the r-process, the rise in [Ba$_{r-process}$/Fe] indicates a significant
increase in production of {\em r}-process elements
near a metallicity [Fe/H]=--2.9.  
The delay in the production until this metallicity may be related to an
evolutionary timescale for the stellar source of the {\em r}-process,
suggesting that most {\em r}-process elements are produced in
relatively low mass (e.g. 8-10~\Msun) Type II supernovae
(Mathews \& Cowan 1990; Wheeler et al. 1998; Ishimaru \& Wanajo 1999;
Travaglio et al. 1999).
Such stars don't evolve quickly enough to contribute {\em r}-process elements
at earlier time (presumably even at the lowest metallicities),
producing a discontinuity when they do begin to contribute.
The transition in [Ba/Fe]$_{r-process}$ at [Fe/H]=--2.9 is qualitatively
different than the gradual {\em s}-process only transition at [Fe/H]=--2.4.
Despite some dispersion, {\rm r}-process element abundances (with
respect to metallicity) rise relatively sharply, indicating
a somewhat narrow range of stellar masses for r-process production.
These {\em r}-process abundances in extremely metal-poor stars
are consistent with the other studies
listed above that suggest low-mass supernovae as the site for the
production of the heavier r-process elements.
Our data alone, however, are insufficient to determine the precise
mass range (e.g. whether 8-10~\Msun stars are responsible) for
such sites of r-process production.

The presence of Ba in stars more metal-poor than [Fe/H]=--3.0 suggests either
contributions from an
additional, earlier (i.e. more massive than 8-10~\Msun) source for the
heavier r-process elements or perhaps a range of r-process/Fe yields
in more massive SN~II.
We note, however, that other studies suggest that such a range
of yields must be relatively small, and likewise any alternative
r-process sources must have conditions somewhat similar to those
found in the standard r-process for the production of the heavier
{\em n}-capture elements. This follows since the abundances of the heavier
{\em n}-capture elements in ultra-metal-poor stars, such as CS 22892-052
([Fe/H]=--3.1), show a scaled solar system r-process pattern.
This close adherence to the Solar System {\em r}-process pattern
implies a relatively narrow range of either nuclear or astrophysical
conditions, and thus constrains any alternative r-process production or
range of r-process/Fe yields to explain the {\em n}-capture abundances
in the early Galaxy.
Thus, for example, the Ba may be produced by a few rare, more massive
Type II supernovae or perhaps by some other process which operates in stars
with masses greater than 10 solar masses.
Note that these more massive supernovae, which may contribute only modest
amounts of {\em r}-process elements, do produce Fe.

Ryan et al. (1996) provide Ba abundances
for a few stars as metal-poor as [Fe/H]=--3.6, which are in agreement
with similar metallicity stars in the McWilliam et al. sample.
These stars are plotted in Figure~7 as crosses to demonstrate one additional
point: that the large scatter seen in the [Ba/Fe]$_{r-process}$ abundances
at metallicities [Fe/H]$>$--2.75 is gone at lower metallicity.
While Ba abundances are available for only a few stars (eliminating
CH/CN-strong stars which may be contaminated from a companion),
the RMS scatter of stars from --3.6$<$[Fe/H]$<$--3.3 is comparable
to the scatter observed in stars near solar metallicity.
To the extent that this reduction in scatter is not the result
of observing Ba only in stars which are {\em n}-capture rich,
the low scatter prior to the onset of the bulk of the
{\em r}-process production at [Fe/H]=--2.9 may constrain the process
producing {\em n}-capture elements at ultra-low metallicity.

Two stars at [Fe/H]=--4.0 in the lower panel of Figure~7 deserve also further 
comment.  These stars, CD~--38~245 and CS~22949-037, have apparent Ba
abundances higher by 0.5~dex than similar giants at [Fe/H]=--3.5.
(We note that a spectrum from Norris (1999) confirms the relatively
high Ba abundance in CD~-38~245.)
Do these giants foretell a rise in the [{\em r}-process/Fe] ratio at
still lower metallicity?
Additional observations of Ba in other stars at [Fe/H]=--4.0 are needed
determine the behavior of the {\rm r}-process ratio at extremely
low metallicity.

\subsection{Lighter N-capture elements: Sr, Y, and Zr}

While the abundances of the heavier (Z$\ge$56) {\em n}-capture
elements can be understood with the {\em r}-process early in the history
of the Galaxy and with {\em s}-process contributions (to at least Ba) 
occurring at higher metallicities and later times, the origin of
the lighter {\em n}-capture elements is not as easily explained.
Unfortunately, the only elements
below Ba that have so far been accessible are Sr, Y and Zr. 
In Figure~8 we illustrate the behavior of Y and Ba with respect to Sr.
In the top panel we show log~$\epsilon$(Ba) vs. log~$\epsilon$(Sr). 
The data sets shown are the same as in Figure~5. Superimposed on the data 
are the scaled Solar System {\em r}-process abundances (dotted line),
the scaled total Solar System abundances (solid line) and
an assumed composition of 10\% {\em r}-process and 90\% weak {\em s}-process
production (dashed line).  The weak {\em s}-process yields have been
taken from Raiteri et al. (1993).
In Figure~9, which plots [Sr/Ba] vs. [Fe/H],
we see a very large scatter in the Sr/Ba abundance ratio
for the lowest metallicity ([Fe/H]$<$--2.5) stars.  While some of this
scatter may be due to observational uncertainties in the Sr abundance,
an additional nucleosynthesis source is required to explain 
the behavior of Sr and the increase in the Sr/Ba ratio at the lowest
metallicities.

The data for the lowest metallicity stars in Figures~8 and 9
do not fall on the Solar System {\em r}-process-only line.
In contrast to the heavier {\em n}-capture elements, the {\em r}-process
alone cannot account for the observed Sr abundances.  In Solar System
material, a sizable contribution ($\sim$20\%) to the Sr abundance
(and less for Y and Zr) comes from the weak {\em s}-process which is
expected to occur much earlier in the proto-Galaxy than the main
{\it s}-process component, since it is produced in core He-burning in
massive ($\sim$ 25 M$_{\odot}$) stars rather than He-shell flashes in
low-mass stars (see Lamb et al. 1977; Raiteri et al. 1993). 
The enhancement of Sr relative to Ba seems to indicate that
there may be an additional {\em n}-capture contribution
to the early synthesis of Sr (in Figures~8 and 9)
with many of the lowest metallicity stars falling on or near the dashed
line representing a composition of 10\% {\em r}-process and 
90\% weak {\em s}-process (see also Ryan et al. 1996).
Not all of the data, however, fall on that line; some, in fact, fall on the
{\em r}-process-only line, suggesting a range of {\em r}-process and
alternate {\em n}-capture contributions, in various amounts, may be required
to explain the Sr abundances in the lowest metallicity stars. 
An additional {\em n}-capture contribution, synthesized in massive stars,
to the abundance of Sr at low metallicities may also add to the
observed scatter in the Sr/Ba ratios in an inhomogeneously mixed proto-Galaxy,
with progenitor stars of perhaps larger masses than those for the 
{\em r}-process production. (This assumes low-mass,  8-10 \Msun, Type~II
SNe progenitors for the {\em r}-process - c.f. Mathews et al. 1992 and
Wheeler et al. 1998.)
While the weak {\em s}-process may be responsible for the enhanced
Sr, Y, and Zr at [Fe/H]$<$--3, we caution that at these very low metallicities,
contributions from the weak {\em s}-process may be too small, due to the
low initial iron abundances in the progenitor stars and the secondary
nature of this process.
(This does not affect {\em r}-process nucleosynthesis.)
On the other hand, the abundances of Sr and Ba
are deficient by typically an order of magnitude relative to Fe at this
low metallicity.

As in Figure~6, we see a transition region shown by the Sr and Ba data in
the upper panel of Figure~8.  The shift toward the Solar System total
line with increasing Sr abundance again indicates an increase in
Ba production at later Galactic time.  This shift, though broad, occurs
near log $\epsilon$(Sr) = 0.6--0.8, corresponding to [Fe/H]=--2.2 to --2.4, 
similar to the range seen for the Ba and Eu data (see also Figure~9). 
Again, this appears to be the onset of the Galactic main {\em s}-process
production affecting the Ba/Sr ratio. 
(The main {\em s}-process, coming from $\simeq$1-3~\Msun\ stars is responsible
for all of the {\em s}-process Ba production - there is no weak
{\em s}-process contribution to Ba synthesis.)
Although there is still some scatter, most of the more metal-rich stars
fall along the total scaled Solar System abundance (solid) line, as expected.  

The lower panel in Figure~8 shows the behavior of Y, also primarily
an {\em s}-process element, with respect to Sr.
In general Y and Sr are better correlated than Ba and Sr.
At lowest metallicities there is again some scatter, but much less
so than in the Ba vs. Sr plot in the top panel.  While the Y and Sr
abundances adhere more closely to the scaled total Solar System
abundance ratio, the Ba and Sr abundances in the upper panel establish
that the main {\em s}-process cannot
be responsible for Sr and Y production at
the lowest metallicities included in Figure~8.  The Sr and Y abundances
are inconsistent with the {\em r}-process being the main source of these
elements in the lowest metallicity stars.
While the weak {\em s}-process makes a more sizable contribution to Sr, Y also
has a weak {\em s}-process component.  A mixture of approximately 80-90\%
weak {\em s}-process and 10-20\% {\em r}-process material is a good match
to the observed Y and Sr abundances in the lowest metallicity stars.
Both the weak {\em s}-process line and the total Solar System (solid) line are 
very close together, and at higher metallicities the data appear to follow 
the total Solar System line, consistent with the dominant
production being by the main {\em s}-process for those higher values of [Fe/H]. 
Observations of the Zr abundance at lower metallicities would be useful
to confirm the contributions of any alternate {\em n}-capture process
in the early Galaxy.

Other alternative {\em n}-capture mechanisms, in addition to the
weak {\em s}-process, have also been suggested to account for the
abundances of Sr, Y, and Zr.
Based upon Solar System meteorite data, a second, separate {\em r}-process
site has been suggested, with one {\em r}-process site for those elements
above Ba and a different site for the lighter {\em n}-capture elements 
(Wasserburg, Busso, \& Gallino 1996; Qian, Vogel, \& Wasserburg 1998).
In addition to massive stars (Woosley et al. 1994) and low-mass supernovae
(Mathews et al. 1992; Wheeler et al. 1998), neutron-star binaries
have also been suggested as possible {\em r}-process sites
(Lattimer et al. 1977; Mathews et al. 1992; Rosswog et al. 1999). 
Examination of our data, particularly Figures~5--10, does seem to 
indicate a different origin for the elements Sr-Y-Zr than
for the heavier {\em n}-capture elements.
We can speculate that the presence of {\em r}-process material early in the 
history of the Galaxy, at very low metallicity, might suggest supernovae
as a likely source for the heaviest {\em n}-capture elements, with (perhaps)
binaries contributing at later times and more infrequently to the
lighter {\em n}-capture abundances.
We caution at this point that the stellar data for the lighter {\em n}-capture
elements, and especially the elements between Zr and Ba, are still not
extensive enough to lead to an unambiguous explanation for the origin of
these elements, and more abundance data and theoretical work will be required. 

\subsection{Galactic Chemical Evolution}

We see further evidence of the extent and nature of the scatter
in the abundances as a function of metallicity in 
Figure~10. This figure plots the geometric means 
($<$[m/Fe]$>$  computed in log space) for the elements Sr-Y-Zr, Ba-La, and
Eu-Dy. The stars have been binned in steps of 0.3 dex, except for
the lowest metallicity range ([Fe/H]$<$--3.6) for which there are
very few stars.  The Eu and Dy abundances for stars more metal-poor
than [Fe/H]$<$--3.0 have been omitted.  These elements have been
measured in only two such metal poor stars (CS-22892-052 and CS-22952-015),
with a few stars showing only upper limits.
(A selection bias may favor {\em r}-process-rich stars.)
The RMS error bars 
(computed in number space) are shown for each group of elements.
Note that the lowest metallicity bin at [Fe/H]=--4.0 contains only
two stars.

Figure~10 illustrates the increase in scatter for {\it all} of 
the elements as metallicity decreases to [Fe/H]=--3.0, supporting
our earlier contention that at early times (i.e., metallicity) the
proto-Galaxy was not chemically well-mixed. The data support the suggestion
of individual nucleosynthetic effects (e.g., supernovae) affecting
different regions of the proto-Galaxy with different total
amounts of synthesized material.  At even lower metallicity,
the scatter in Sr-Y-Zr (mainly Sr at this low metallicity) continues
to increase, while the scatter in Ba-La (mainly Ba) seems to decrease,
as mentioned earlier in \S 4.3.

Figure~10 also demonstrates
that the overall scatter seems to have dropped 
to zero, and the mean abundances are roughly consistent with solar values, 
by the time [Fe/H]=--0.6. This presumably marks the 
metallicity (and some associated Galactic time) when the 
stellar chemical abundances have become homogenized
and when the Galaxy became more thoroughly chemically mixed
(see also McWilliam 1998). This may also be related to 
the disk-halo transition. 

At very low metallicities ([Fe/H]$\le$--2.5), the average abundance
relative to Fe
of the elements Sr, Y, Zr, Ba, and La is significantly subsolar, indicating
that the production of many {\em n}-capture elements
lags behind early Galactic iron production (presumably from stars above
10\Msun). As metallicity increases above [Fe/H]$\sim$--2.0, the average
abundances of these heavy elements relative to Fe
may rise slightly above solar values.
For the heavier {\em r}-process elements Eu and Dy, the abundances are
markedly super-solar, and they show no evidence for the decline characteristic
of the ``{\em s}-process'' elements at [Fe/H]$<$--2.5.
Later, presumably with the increasing iron production from Type~I supernovae,
beginning at [Fe/H]$\sim$--1.5, the abundance ratios tend downward toward
solar values, suggesting that iron production is not tied to the
same astrophysical site as {\it r}-process nucleosynthesis.
Instead, it may be that supernovae from only a narrow mass range of
perhaps 8--10 \Msun\ (i.e., low-mass) stars 
are responsible for {\it r}-process nucleosynthesis
(Mathews \& Cowan 1990; Wheeler et al. 1998;
Ishimaru \& Wanajo 1999; Travaglio et al. 1999),
while early iron synthesis is coming from a different range of massive stars
producing Type II supernovae. 
Furthermore, if the neutron-capture 
elements and iron are both coming from Type II supernovae,
our data implies that the [{\em r}-process/Fe] ratio of the ejecta
must exceed the solar ratio, and that the {\em r}-process abundance
is increasing at a faster rate than the Fe abundance early during
the epoch after the first 8-10 \Msun\ stars become supernovae and
before the SN~Ia's begin to contribute Fe.
Alternatively, other possible {\em r}-process sites
might also participate in this early Galactic
{\it r}-process nucleosynthesis, and would not be involved in iron synthesis.
One such suggested site, for example,
is neutron star-binaries (see Mathews et al. 1992; Rosswog et al. 1999).
(See also Cowan et al. 1991 for a  
discussion of possible {\em r}-process sites.)

Figure~10 also demonstrates a difference in the behavior between
the lighter (Sr-Y-Zr) and the heavier (Ba-La and Eu-Dy)
{\em n}-capture elements.
For example, the abundances of Sr, Y, and Zr are significantly
higher than Ba and La at very low metallicities, as expected if an
additional production mechanism occurring in massive stars is involved.
As we have noted earlier, at metallicities above --2.75, both Sr, Y, and Zr
and Ba and La are
produced in the main {\em s}-process, while several of the heavier elements,
notably Eu and Dy are produced primarily in the {\em r}-process.
The difference in the mean values may reflect the evolution
of the different (mass) sites for these synthesis processes.
Some contribution to the scatter in the lighter element data
at metallicities [Fe/H]$<$--2 may be from the weak
{\em s}-process coming from massive stars.

\subsection{Abundances at Very Low ([Fe/H]$<$--3.5) Metallicities}

We end this discussion of the abundances by stressing 
the need for additional data for stars
with [Fe/H]$<$--3 to draw any conclusions on 
the nature of the early synthesis of these {\em n}-capture elements.
Thus, for example, if Ba and Fe are both produced 
(early in the Galaxy) in Type~II supernovae, can we
expect that the [Ba/Fe] abundance ratio should remain constant at
[Ba/Fe]=--1.3 at lower metallicity (e.g., --3.5$<$[Fe/H]$<$--4.5)?
Should the [Ba/Eu] ratio also remain constant at [Ba/Eu]=--0.89?
Observations of [Ba/Fe] and [Eu/Fe] at very low metallicity,
would also help to confirm whether low-mass (8-10~\Msun) Type~II
supernovae are the dominant site for the {\em r}-process, and to
identify the site of any {\em r}-process nucleosynthesis at earlier
Galactic time.
Such observations may be difficult, however, considering the weakness of
the Eu II lines in most giants already at [Fe/H]=-3.0, and of even the Ba II
lines at [Fe/H]=--4.0.
Observations are also needed at very low metallicities for Y and Zr to see if 
they mimic the trends seen (and presumably the production mechanisms) for Sr.
Finally, we note that observations in very low-metal stars of elements between
Zr and Ba could help in deciding whether there are, in fact,
two separate {\em r}-process sites and if the production mechanisms in 
those sites were the same in the early Galaxy as they may be today.

\section{CONCLUSIONS}

We have presented new {\em n}-capture abundance results for 43 metal-poor Bond
giants and Ba abundances only for an additional 27 metal-poor giants.  From
these heavy element abundances, we draw five principal conclusions.

\noindent
1.  {\it The scatter in the abundances of all the {\em n}-capture elements
from star to star is of astrophysical origin, and the 
scatter increases as we go to lower metallicities.}
Elemental abundance correlations (e.g., [Dy/Eu] vs. [Fe/H])
demonstrate that all of the observed heavy elements with Z$\ge$56
vary together, and the observed scatter is not the result of errors
in measurement or analysis.  Instead, this scatter results from
inhomogeneity of the material at early times in the Galaxy's history, 
presumably because of the mixture of different amounts of supernova
nucleosynthesis ejecta products in the gas at different star formation sites.
While this result was first noted by Gilroy et al. (1988) and was 
emphasized by McWilliam et al. (1995b), 
our new comprehensive data, covering a wider metallicity range,
strengthen this contention.  As the metallicity increases, and 
presumably the age of the Galaxy, the scatter decreases, indicating that
production from more common sites and/or
mixing of the material must be taking place.  

\noindent
2.  {\it The ratio of the abundances of the elements Ba (an {\em
s-}process element in the Sun) and Eu (an {\em r-}process element in the Sun)
at very low metallicity indicates that the heavy {\em n}-capture elements
were formed predominantly via the {\em r-}process at the earliest
times in Galactic history.}
The ratio of these elements in the most metal-poor stars is reduced by a
factor of six compared to the solar composition, and the ratio 
remains at or near the solar {\em r}-process value even as the
overall abundance of the {\em n}-capture elements relative
to iron varies by more than an order of magnitude.
This result, suggested previously based upon a single or a 
few stars (c.f. Truran 1981; McWilliam et al. 1995b;
Sneden et al. 1996, Ryan et al. 1996) is now seen to be
characteristic of the halo population in the Galaxy.
For these very low-metallicity stars, the relative {\em n}-capture elemental
abundances are consistent with a scaled-Solar System {\it r}-process
distribution, at least for elements with Z $\ge$ 56.
These new results put significant constraints on the types 
(and specifically the evolutionary timescales) of 
objects that were the progenitors of the metal-poor halo stars and
on the nucleosynthetic conditions that produced the {\em r}-process elements.

\noindent
3.  {\it Significant production of {\em r}-process elements
began when the metallicity of the Galaxy reached [Fe/H]=--2.9.}
This onset is consistent with
the original suggestion by Mathews et al. (1992) that Type II supernovae
of a narrow range of initial mass (8-10 \Msun) stars are responsible
for nucleosynthesis of {\rm r}-process elements.
Some {\rm r}-process element production also occurred to produce the
{\em r}-process elements observed at even lower metallicity, perhaps
due to rare, more massive SN~II's or to other processes operating
in such massive stars.  
Observations of Ba in stars with metallicities
near [Fe/H]=--4.0 are needed to understand such early nucleosynthesis.

\noindent
4. 
{\it Our abundance correlations also suggest that an alternate {\em n}-capture
site is needed for much of the production of the lightest {\em n}-capture
elements, specifically Sr, early in the history of the Galaxy.}  For
the most metal-poor stars, the elements Sr, Y, and Zr do not
follow the same trends as the heavier {\em n}-capture elements,
indicating that Sr-Y-Zr are probably not formed via the same process
as the elements such as Ba, Eu, etc.
Sr does correlate with Ba at metallicities above [Fe/H] = --2.5 where
both elements are produced primarily in the main {\em s}-process 
in low-mass AGB stars.  For
lower metallicities, the Sr/Ba ratio increases and the stars display
a large scatter in the Sr/Ba ratio.  The lack of a correlation between the Sr and 
Ba abundances at very low metallicity suggests that different processes
are responsible for their formation.  For [Fe/H]$<$--2.8 a combination
of the {\em r}- and perhaps the weak {\em s}-process 
could be responsible for light element {\em n}-capture production.
At [Fe/H]=--3 the abundances of Sr and Y are consistent (on average) with
a mixture of 80-90\% contributions from the weak {\em s}-process and
10-20\% contributions from the {\em r}-process.
Note that a correlation of Sr with Ba would not be expected, since the weak
{\em s}-process produces Sr but not Ba or other, heavier, {\em n}-capture elements.
A second {\em r}-process may also have contributed to the Sr, Y, Zr, and
other light (Z$<$56) {\em n}-capture element abundances. 
In order to clarify the processes which produce the light {\em n}-capture elements
in the most metal-poor stars, observations of additional elements lying between
Zr and Ba will be needed.  Such elements may be produced by
the second {\rm r}-process but not the weak {\em s}-process.

\noindent
5. 
{\it As metallicity
increases, {\em s}-process contributions from low-mass AGB stars become
apparent and finally dominate the Ba abundance, leading to the observed
Solar System ratio of Ba/Eu.}
The data indicate the presence of a main {\em s}-process contribution
in some stars as metal-poor as [Fe/H]=--2.75, lower than previous studies
have suggested, with the major increase in (the bulk of the) Ba production
occurring at [Fe/H]=--2.2 $\pm$ 0.2. 
The transition to an {\em s}-process regime is delineated clearly in our
data because of the large sample of stars now available.  The transition
is difficult to pinpoint without a large sample because of the scatter
in [Ba/Fe] at low metallicities and the spread of metallicity over which
the transition occurs.  The observed decline of Ba relative to Fe
at low metallicity is due to the absence of {\em s}-process contributions
to the heavy element abundances below [Fe/H]=--2.8, rather than because
Ba is a secondary element.  The delay in the onset of the {\em s}-process
is consistent with standard models suggesting lower masses (and hence
longer evolutionary timescales) for the sites of the {\em s}-process.
However, the onset of the {\em s}-process in low-mass AGB stars near
[Fe/H]=--2.8 along with a range of metallicity over which Galactic
{\em s}-processing occurs, suggests both a range of {\em s}-process progenitors and
that the earliest occurrence of Galactic {\em s}-processing might arise from 
progenitors more massive than typically assumed for the main {\em s}-process.

\acknowledgments

We thank Daryl Willmarth and the staff of the Kitt Peak National
Observatory for assistance in obtaining the spectra used in our
analysis, and David Arnett, Dave De Young, Inese Ivans,and Craig Wheeler
for valuable conversations
on the subject of the early chemical enrichment of the Galaxy,
and Roberto Gallino for his extensive and thoughtful comments
on the manuscript.
We also thank an anonymous referee for thoughtful advice which
significantly improved this manuscript.
Partial support for this research was provided by 
the National Science Foundation 
(AST-9618364 to C.S. and AST-9618332 to J.J.C.),    

\newpage
\begin{center}
{\bf Appendix A\\
     Updated Solar Elemental Abundances}
\end{center}
Using the updated cross-sections of Wisshak, Voss, \& K\"appeler (1996) , 
we have recalculated the Solar System {\em s-} and {\em r-}process 
fractions for each element.  
Since only a portion of the isotopes in this range have been reanalyzed 
by Wisshak et al., we return to K\"appeler, Beer, \& Wisshak (1989) 
for the remainder of the cross-sections.
Beginning with the total Solar System abundances of Anders \& Grevesse 
(1989), the individual isotopic contributions to each element were 
obtained from the Chart of the Nuclides.  
We then separated out the {\em s-} and {\em r-}process contributions 
to each isotope taken from either K\"appeler et al. or Wisshak et al. 
as appropriate and summed over all the isotopes for each process.
In Table~4, we list the {\em s-} and {\em r }process
elemental abundances computed in this manner.  
This table is similar to Table~5 in Sneden et al. 1996, with corrections 
to errors for Rb and Xe.  
In addition, we have included abundances for Lu and U, which were
not included in our original table.
In columns~3-5 we include the abundances for {\em r-}process, 
{\em s-}process, and total solar abundances.  
In columns~6 and 7 we list the fraction of total for the 
{\em r-}process ($\equiv$F$_\odot$[r]) and the {\em s-}process
($\equiv$F$_\odot$[s]), respectively.
In Figure~11 these {\em r-} and {\em s-}process fractions are displayed
so that one may easily note the dominant nucleosynthesis contribution 
to each element in the solar system distribution.
In the figure, atomic symbols are entered near the {\em r-}process fractional
point for those elements with solar {\em r-}fractions F$_\odot$[r]$>$0.75
or solar system {\em s-}fractions F$_\odot$[s]$>$0.75.
Thus for example the solar system {\em s-}process-dominated element Ce
(F$_\odot$[s]=0.8142, Table~4) is labeled, as is the
{\em r-}process-dominated element Ir (F$_\odot$[r]=0.9924),
but the mixed-process element Nd (F$_\odot$[r]=0.5291, F$_\odot$[s]=0.4709) 
is unlabeled.

\begin{table}
\dummytable\label{tab-abunds}
\end{table}

\clearpage
\pagestyle{empty}

\noindent{\bf Figure Captions:}

\vskip .2truein
\figcaption{Top panel: Comparison of iron abundance from this work with PSK.
lower panel: comparison of [Ti/H] with [Ca/H] from PSK.
\label{fig1}}
\vskip0.25cm

\figcaption{ Spectrum synthesis for (a) (lower panel) the  La 4123 \AA 
\ and  (b) (upper panel) Eu 4205 \AA  \ lines for HD 63791.
\label{fig2}}
\vskip0.25cm

\figcaption{Spectral comparison at three wavelengths between two metal-poor stars
with similar metallicities and atmospheric parameters, BD~+9~2870 (open circles) and
HD~6268 (solid line).  Note the similarity of Fe-peak lines in the upper panel
and the difference in strength in the {\em n}-capture lines in the lower two panels.
\label{fig3}}
\vskip0.25cm

\figcaption{Abundance comparisons  of (a, top panel)  [Eu/Fe], 
(b, middle panel) [Dy/Fe] and (c, bottom panel) [Dy/Eu]
in metal-poor stars. HD 6268 is indicated by an open square and BD~+9~2870 by
an open circle.
The reduction in scatter in the lowest panel compared to the upper panels
reflects the strong correlation between the measured Eu and Dy abundances,
and establishes that the scatter is not the result of observational
uncertainty.
\label{fig4}}
\vskip0.25cm

\figcaption{[Sr/Fe], [Y/Fe], [Zr/Fe],  
[Ba/Fe], [La/Fe], [Nd/Fe], [Eu/Fe] and [Dy/Fe] as a function of metallicity.
Filled circles are from this paper (filled triangles represent
upper limits for Eu and Dy), open triangles
from McWilliam et al. (1995a, 1995b) and McWilliam (1998), 
open squares Jehin et al. (1999), and open circles
from Edvardsson et al. (1993) and Woolf et al. (1995).   Note that the
McWilliam et al. sample contains two known or suspected CH stars, which
have s-process element enrichments due to mass transfer from a more massive
and evolved companion;
these stars have been eliminated from the sample.
\label{fig5}}
\vskip0.25cm

\figcaption{Log $\epsilon$(Eu) versus log $\epsilon$(Ba).
The dotted line indicates the Solar System {\em r}-process abundance ratio
and the solid line indicates the total Solar System abundance ratio of Ba/Eu.
The symbols are as in Figure~5.
\label{fig6}}
\vskip0.25cm

\figcaption{[Ba/Fe] vs. [Fe/H] for the {\em r}-process-only (lower
panel) and {\em s}-process-only portions of Ba.  The {\em r}-process
and {\em s}-process fractions were computed as described in the text.
The symbols are as in Figure~5, with the addition of crosses for stars from
Ryan et al. (1996).  The downwardly directed arrows in the upper panel
mark the position of stars which do not appear to contain s-process material.
\label{fig7}}
\vskip0.25cm

\figcaption{Log $\epsilon$(Ba) (upper panel) and log $\epsilon$(Y) (lower panel)
versus log $\epsilon$(Sr).
The dotted lines in both panels mark the Solar System {\em r}-process abundance ratio,
the solid lines are the total Solar System abundance ratio, and the dashed lines
assume a composition of 90\% weak {\em s}-process + 10\% {\em r}-process. 
The symbols are as in Figure~5.
\label{fig8}}
\vskip0.25cm

\figcaption{[Sr/Ba] versus [Fe/H] for metal-poor giants. 
The dashed line indicates a composition of 90\% weak
{\em s}-process + 10\% {\em r}-process,
the solid line marks the solar ratio and the dotted line is the solar
{\em r}-process-only ratio. Symbols are as in Figure~5.
\label{fig9}}
\vskip0.25cm

\figcaption{
 $<$[m/Fe]$>$ vs [Fe/H] and RMS scatter vs. [Fe/H]  for
Sr-Y-Zr (filled squares and dotted line), Ba-La (filled circles
and solid line) and Eu-Dy (open circles and dashed line).
The data are from this paper, from McWilliam et al. (1995a, 1995b)
and McWilliam (1998), Jehin et al. (1999), Ryan et al. (1996),
and Edvardsson et al. (1993) and Woolf et al. (1995). 
Plotted are the geometric means of the abundances and the RMS scatter
computed in number space.  The [Fe/H] values have been shifted slightly
to avoid overlaying the errorbars.
Note the reduction in the scatter of the Ba-La abundances at [Fe/H]=-3.5.
The bin at [Fe/H]=-4.0 contains only two stars.
\label{fig10}}
\vskip0.25cm

\figcaption{ {\em r}-process and {\em s}-process fractions for Solar
System elemental abundances F$_\odot$[r] and F$_\odot$[s], computed
as described in Appendix A. 
\label{fig11}}

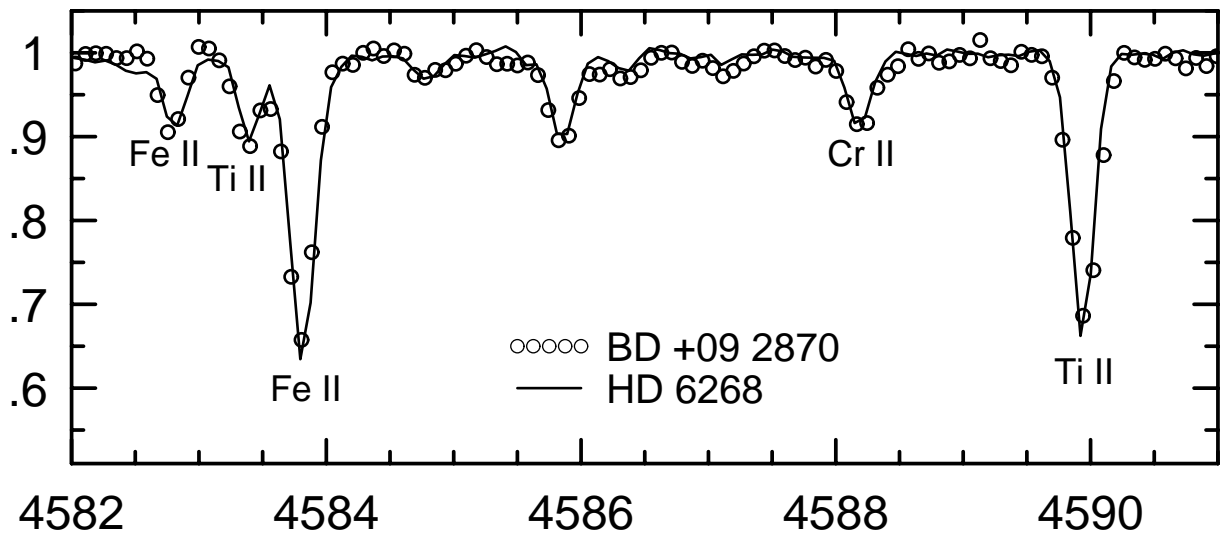
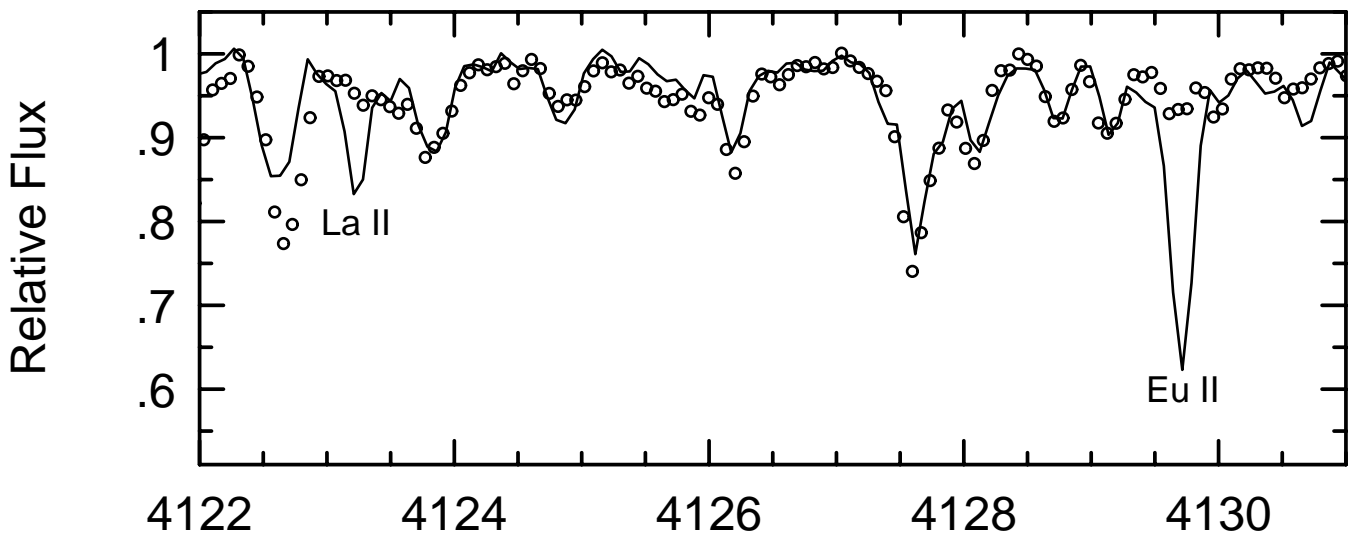
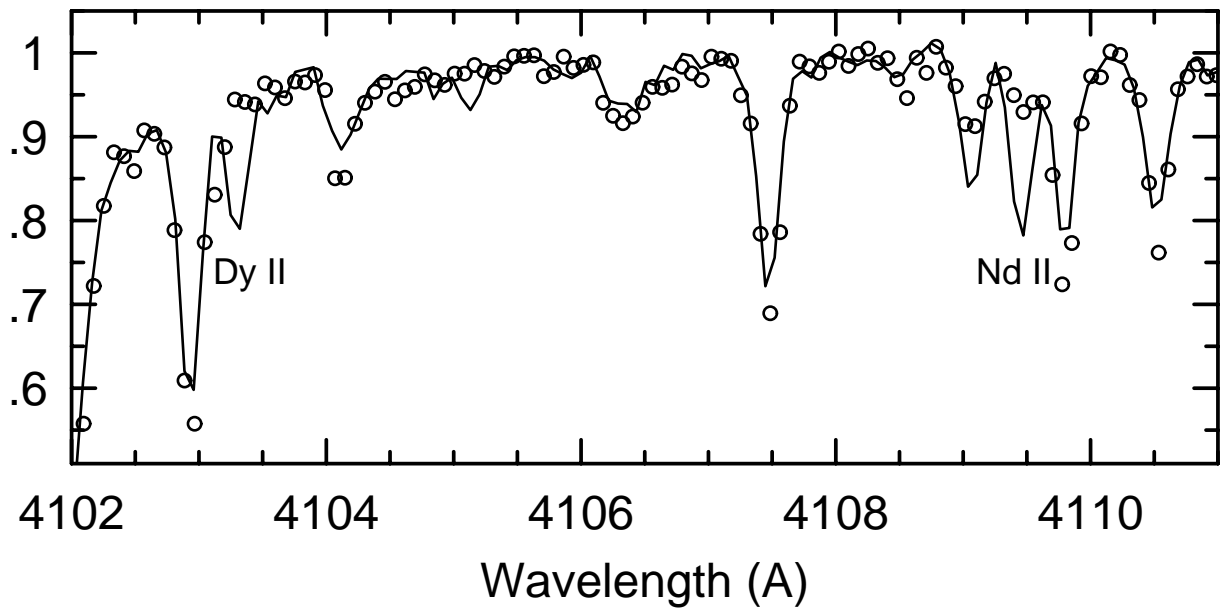
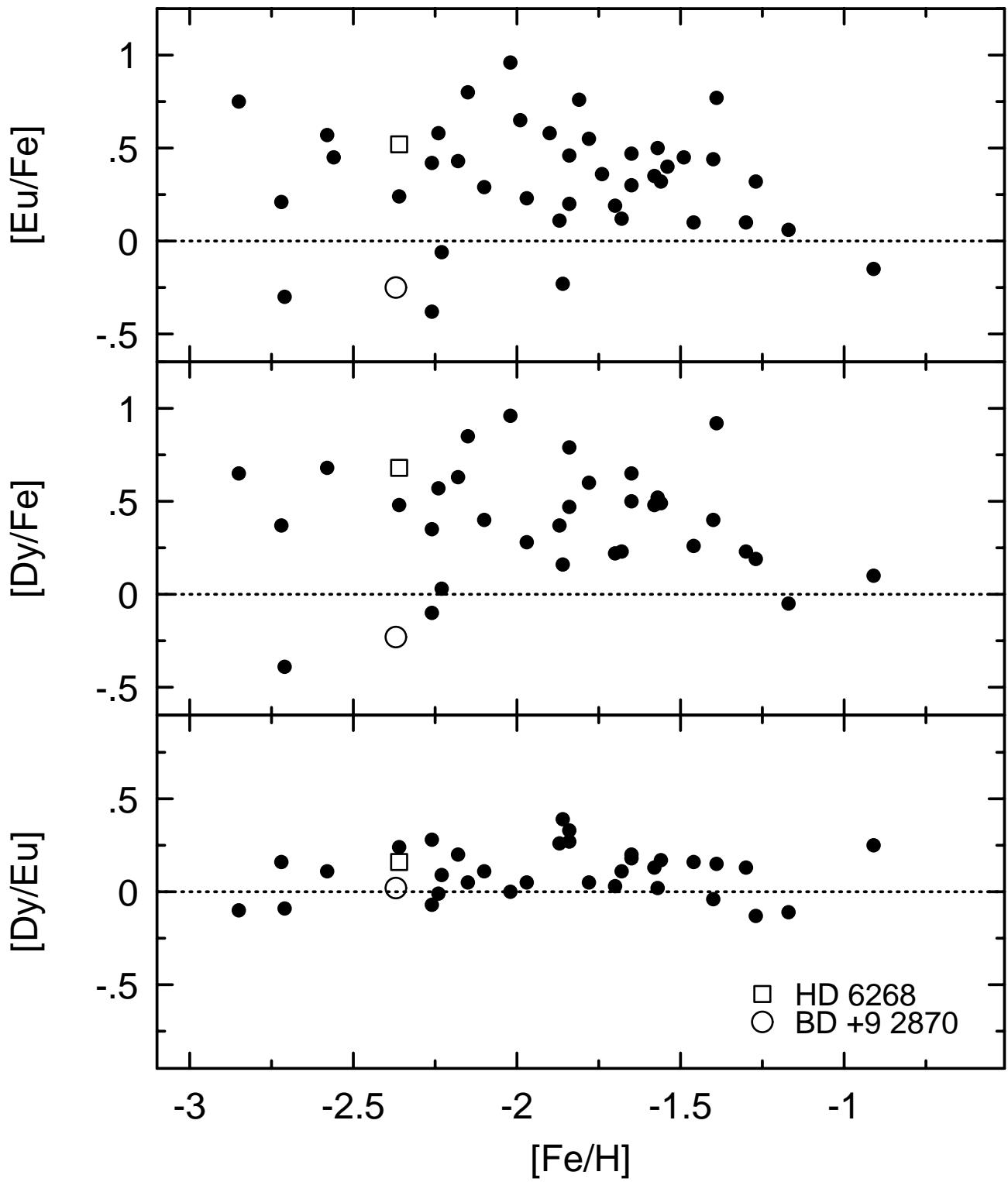
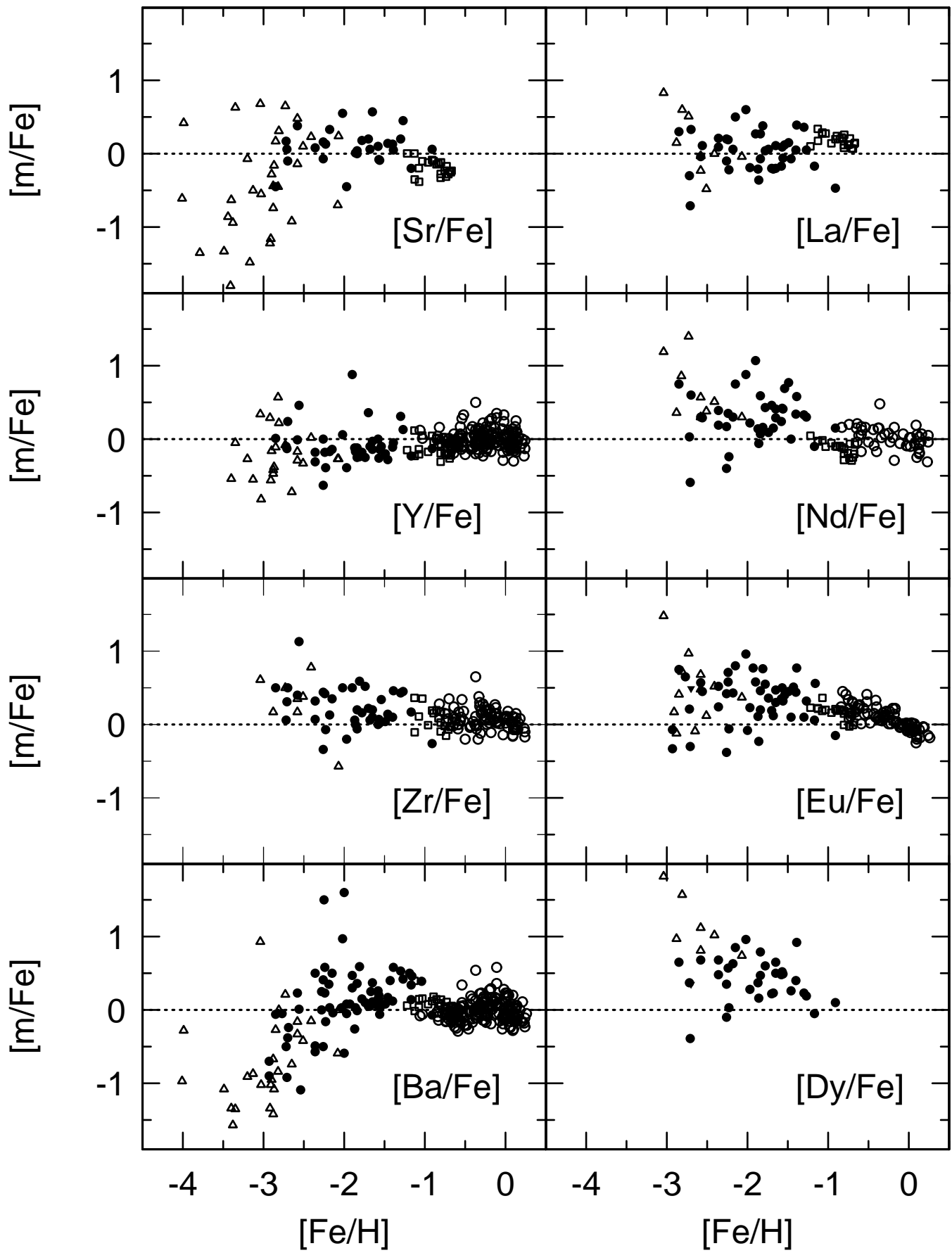
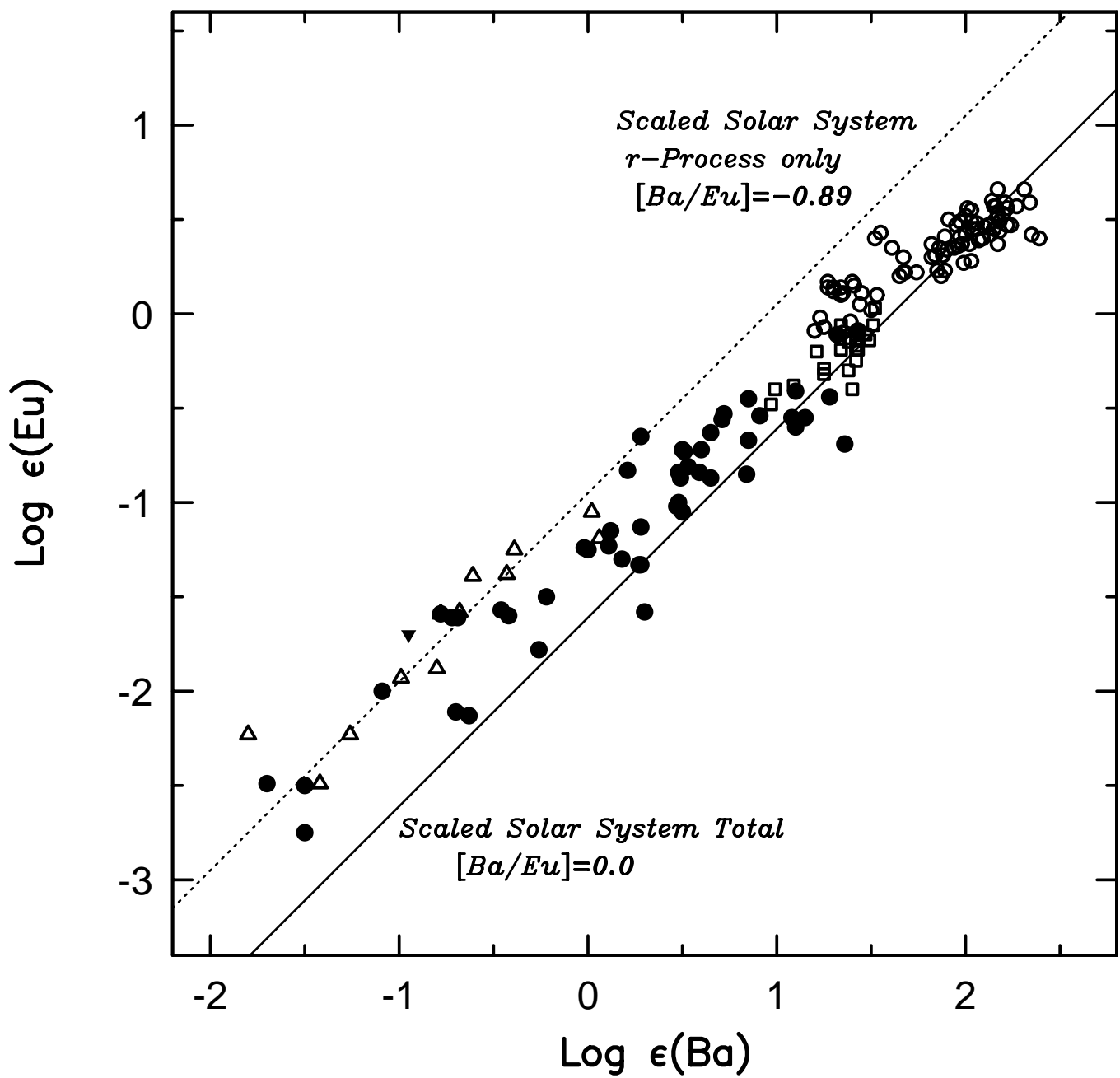
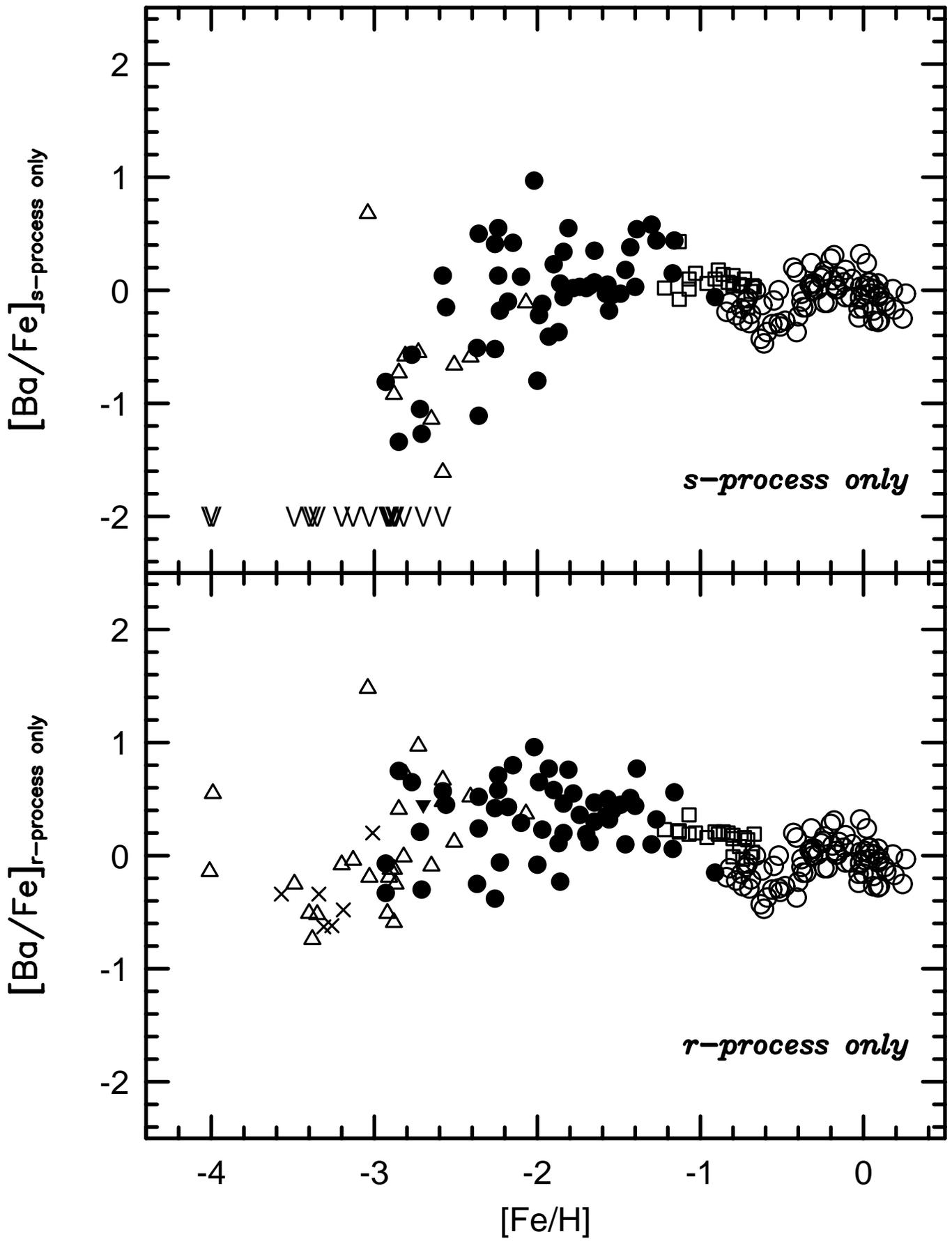
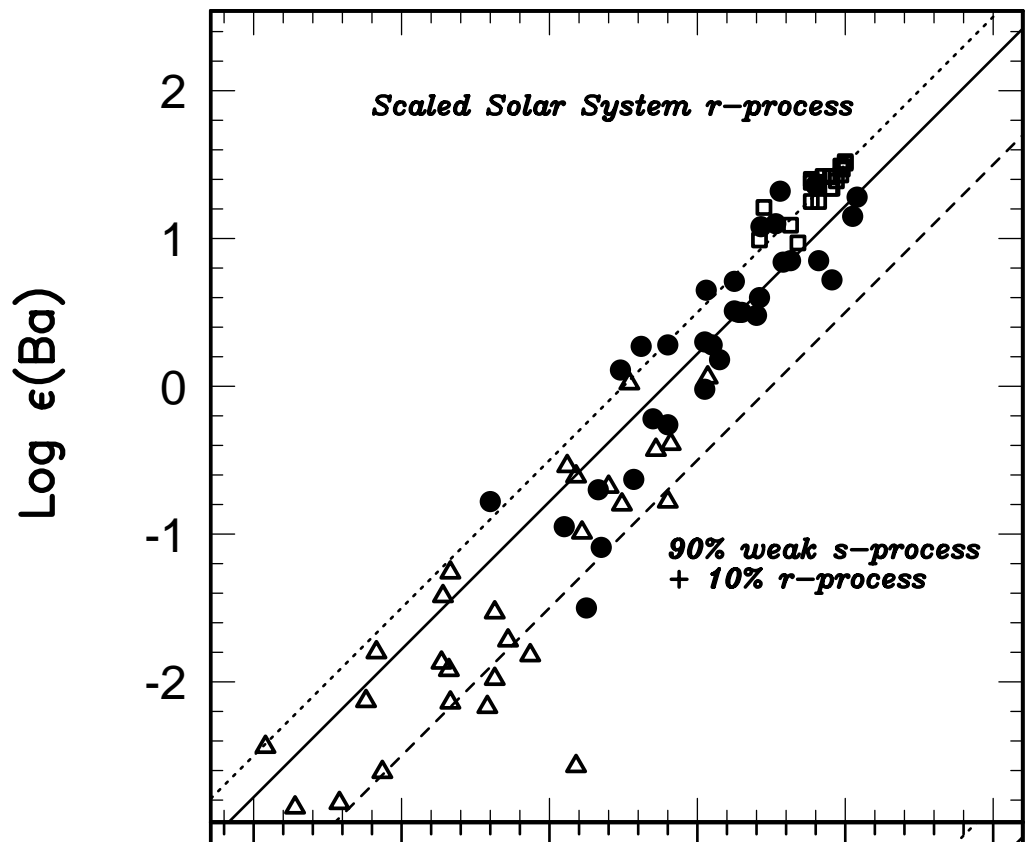
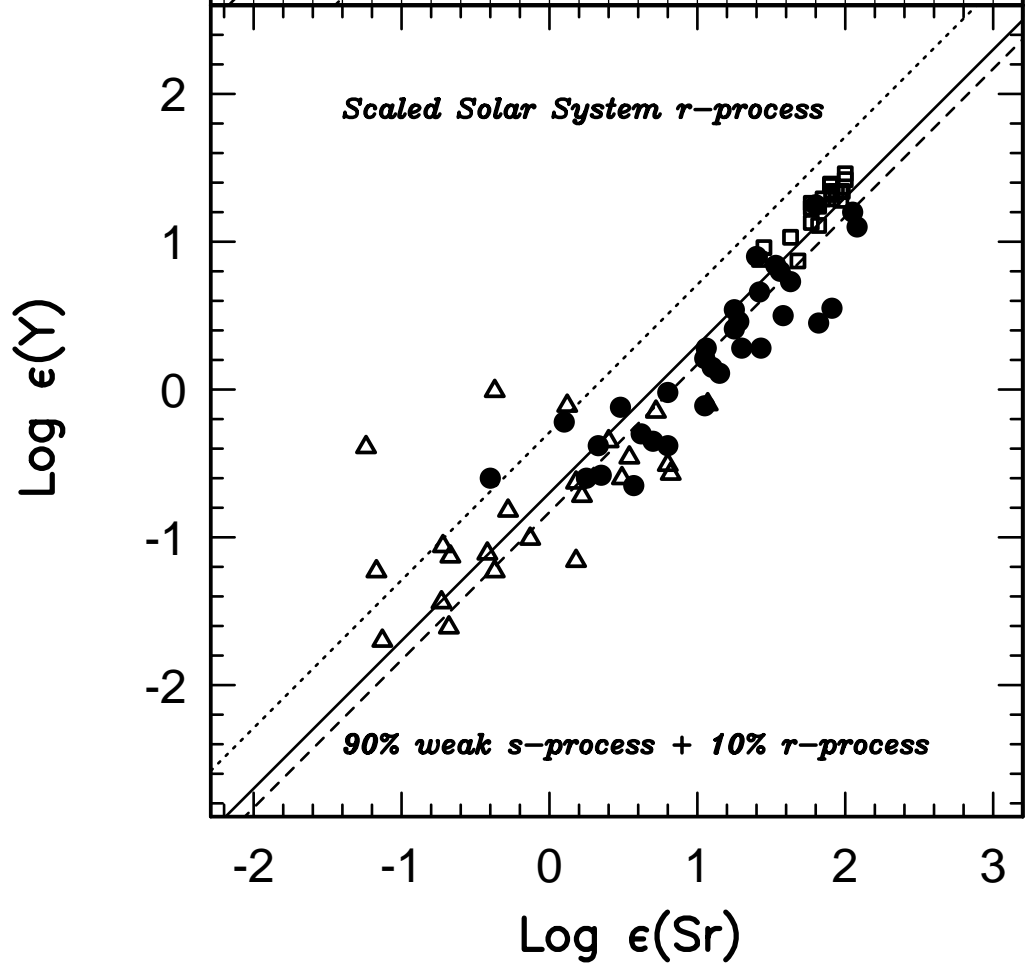
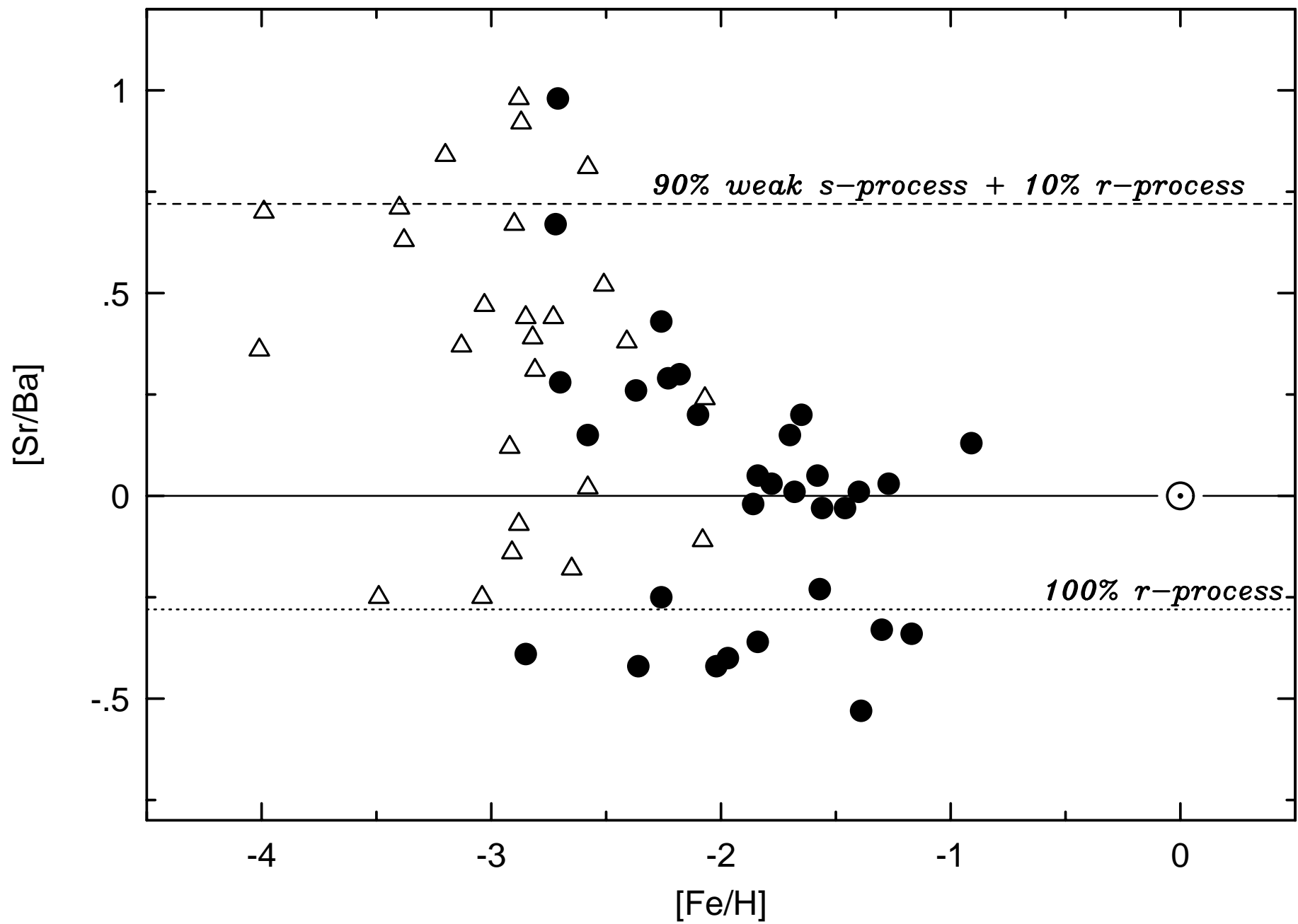
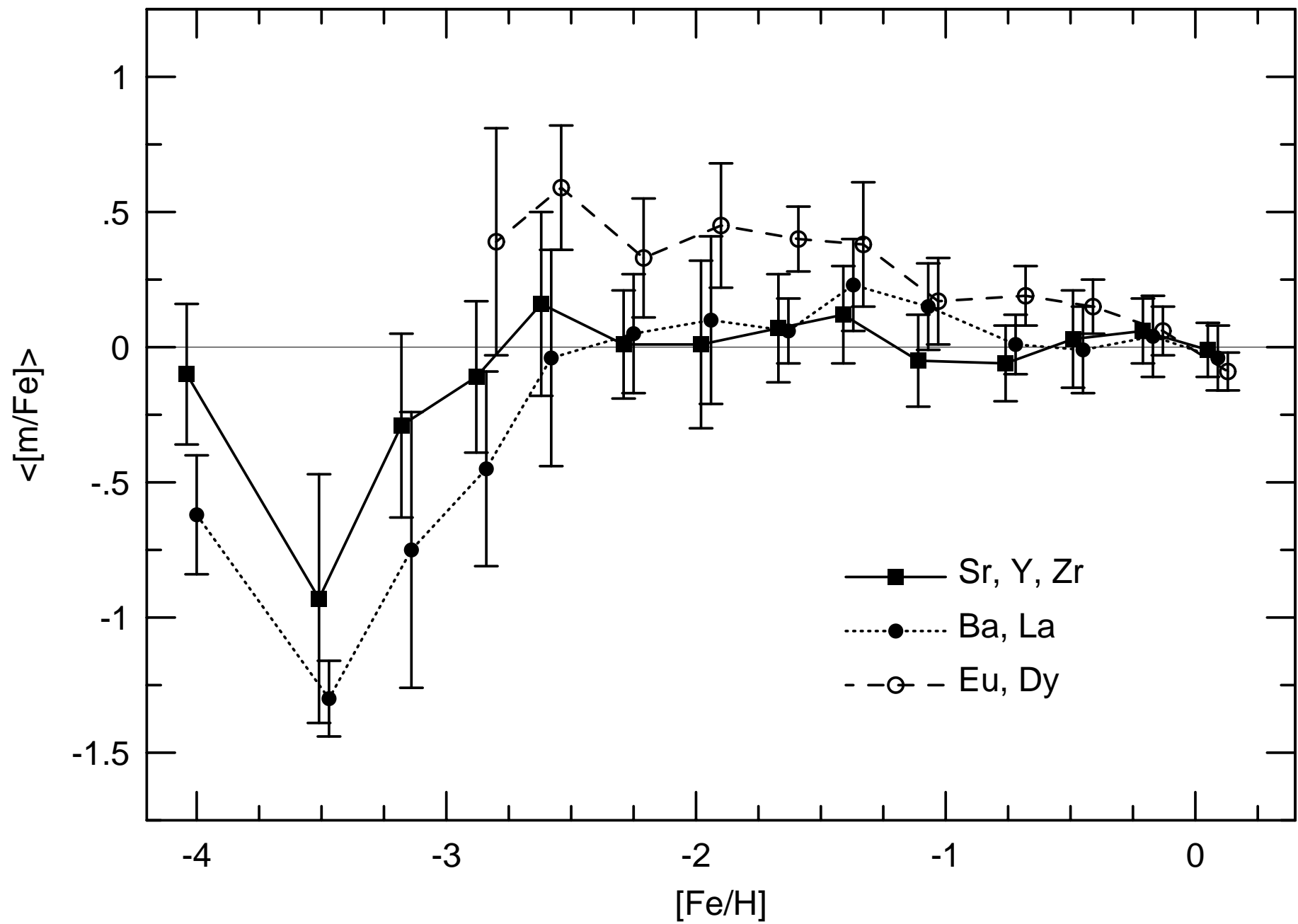
\end{document}


\begin{deluxetable}{lcllllc}
\tablenum{1}
\tablewidth 4.5in
\tablecaption{Model Atmosphere Parameters}
\tablehead{
\colhead{} & \colhead{} & \colhead{T$_{eff}$} & \colhead{} & \colhead{} & \colhead{v$_{t}$} \nl
\colhead{Star}       &
\colhead{Model\tablenotemark{a}}      &
\colhead{(K)}        &
\colhead{log g}      &
\colhead{[M/H]}      &
\colhead{km s$^{-1}$} &
\colhead{Note}
}
\startdata
HD     20   &   PSK &  5475 & 2.80  & $-$1.20 &    2.00 & \nodata \nl
HD     97   &   PSK &  5025 & 2.60  & $-$1.20 &    1.20  &  1 \nl
HD   2665   &   PSK &  5000 & 2.20  & $-$3.00 &    1.30 & \nodata \nl
HD   2796   &   PSK &  4900 & 1.60  & $-$3.00 &    1.60 & \nodata \nl
HD   3008   &   PSK &  4150 & 0.60  & $-$2.00 &    2.20  & 1,2 \nl
HD   3179   &  (PSK)&  5300 & 2.60  & $-$0.80 &    1.90  &  1 \nl
HD   4306   &   PSK &  4900 & 2.00  & $-$3.00 &    1.40 &   1 \nl
HD   6268   &   PSK &  4700 & 1.60  & $-$3.00 &    1.60 & \nodata \nl
HD   6755   &   PSK &  5150 & 2.70  & $-$1.50 &    1.40 & \nodata \nl
HD   8724   &   PSK &  4500 & 1.20  & $-$2.20 &    1.60 & \nodata \nl
HD  13979   &   PSK &  5075 & 1.90  & $-$3.00 &    1.30 & \nodata \nl
HD  25532   &   PSK &  5300 & 1.90  & $-$1.20 &    2.30 & \nodata \nl
HD  26297   &   PSK &  4400 & 1.10  & $-$1.70 &    2.00 & \nodata \nl
HD  29574   &   new &  4250 & 0.00  & $-$1.65 &    1.90 & \nodata \nl
HD  44007   &   PSK &  4850 & 2.00  & $-$1.50 &    1.50 & \nodata \nl
HD  63791   &   PSK &  4725 & 1.70  & $-$1.50 &    1.60 & \nodata \nl
HD  74462   &   PSK &  4600 & 1.50  & $-$1.40 &    1.80 & \nodata \nl
HD  83212   &   PSK &  4550 & 1.50  & $-$1.20 &    1.80  & 1,2 \nl
HD  85773   &   PSK &  4450 & 1.10  & $-$2.00 &    2.10 & \nodata \nl
HD  88609   &   PSK &  4575 & 1.40  & $-$3.00 &    1.80  & 1,2 \nl
HD  93529   &   PSK &  4650 & 1.70  & $-$1.50 &    1.40  &  1 \nl
HD 103545   &   PSK &  4725 & 1.70  & $-$3.00 &    1.30  &  1 \nl
HD 104893   &   PSK &  4500 & 1.10  & $-$2.20 &    1.90  &  1 \nl
HD 105546   &   PSK &  5300 & 2.50  & $-$1.00 &    1.50 & \nodata \nl
HD 107752   &   PSB &  4700 & 1.70  & $-$3.00 &    1.40  &  1 \nl
HD 108317   &   PSK &  5300 & 2.90  & $-$2.20 &    1.00  & 1,2 \nl
HD 108577   &   PSK &  4975 & 1.70  & $-$1.50 &    1.90  &  1 \nl
HD 110184   &   PSK &  4250 & 0.30  & $-$2.20 &    2.10 & \nodata \nl
HD 115444   &   PSK &  4750 & 1.70  & $-$3.00 &    1.60  & 1,2 \nl
HD 121135   &   PSK &  4925 & 1.50  & $-$1.50 &    2.00  &  1 \nl
HD 122563   &   PSK &  4625 & 1.40  & $-$3.00 &    1.70 & \nodata \nl
HD 122956   &   PSK &  4600 & 1.50  & $-$1.70 &    1.80 & \nodata \nl
HD 126587   &   McW &  4910 & 1.85  & $-$3.00 &    2.00 & \nodata \nl
HD 128279   &   PSK &  5275 & 2.80  & $-$2.00 &    1.00  & 1,2 \nl
HD 135148   &   PSK &  4275 & 0.80  & $-$1.70 &    2.10 & \nodata \nl
HD 165195   &   PSK &  4450 & 1.10  & $-$2.00 &    1.90 & \nodata \nl
HD 166161   &   PSK &  5150 & 2.20  & $-$1.00 &    2.00 & \nodata \nl
HD 171496   &   PSK &  4700 & 1.60  & $-$1.00 &    1.50 & \nodata \nl
HD 175305   &   PSK &  5100 & 2.50  & $-$1.50 &    1.40 & \nodata \nl
HD 186478   &   PSK &  4575 & 1.40  & $-$2.20 &    2.00 & \nodata \nl
HD 187111   &   PSK &  4250 & 0.70  & $-$1.70 &    1.70 & \nodata \nl
HD 204543   &   PSK &  4700 & 1.70  & $-$1.50 &    2.00 & \nodata \nl
HD 206739   &   PSK &  4675 & 1.70  & $-$1.70 &    1.70 & \nodata \nl
HD 214362   &   PSK &  5650 & 1.50  & $-$1.70 &    1.80  &  1 \nl
HD 216143   &   PSK &  4525 & 0.80  & $-$2.00 &    1.80 & \nodata \nl
HD 218857   &   PSK &  5125 & 2.40  & $-$2.00 &    1.20 & \nodata \nl
HD 220838   &   PSK &  4450 & 1.20  & $-$1.70 &    1.90 & \nodata \nl
HD 221170   &   PSK &  4425 & 1.00  & $-$2.00 &    1.50 & \nodata \nl
HD 232078   &   PSK &  4000 & 0.30  & $-$1.50 &    2.60 & \nodata \nl
CD $-$24 1782 &   PSB &  5300 & 2.80 & $-$2.20 &    1.50 & \nodata \nl
CD $-$23 72   &   PSK &  5300 & 2.60 & $-$1.00 &    2.00  &  1 \nl
BD $-$18 5550 &   PSK &  4575 & 1.40 & $-$3.00 &    1.80  & 1,2 \nl
BD $-$15 5781 &   PSK &  4750 & 1.20 & $-$3.00 &    1.70  &  1 \nl
BD $-$11 145  &  (PSK)&  4800 & 1.70 & $-$2.00 &    1.60  &  1 \nl
BD $-$3 5215  &   PSK &  5700 & 2.80 & $-$1.20 &    1.80  & 1,2 \nl
BD $-$1 2582  &   PSK &  5150 & 2.50 & $-$2.00 &    1.20  &  1 \nl
BD +1 2916  &   PSK &  4150 & 0.10  & $-$1.80 &    2.30 & \nodata \nl
BD +3 2782  &  SKPL91& 4600 & 1.50  & $-$2.00 &    1.80  &  1 \nl
BD +4 2466  &  (PSK)&  5000 & 1.00  & $-$1.80 &    1.30  &  1 \nl
BD +6 648   &   PSK &  4500 & 1.10  & $-$2.00 &    2.00 & \nodata \nl
BD +9 2870  &   PSK &  4600 & 1.40  & $-$2.20 &    1.70 & \nodata \nl
BD +9 3223  &   PSK &  5350 & 2.00  & $-$2.20 &    2.00 & \nodata \nl
BD +11 2998 &   PSK &  5425 & 2.30  & $-$1.20 &    2.20 & \nodata \nl
BD +17 3248 &   PSK &  5250 & 2.30  & $-$1.80 &    1.50 & \nodata \nl
BD +18 2757 &   PSK &  4850 & 1.50  & $-$1.40 &    1.90  &  1 \nl
BD +18 2890 &   PSK &  5000 & 2.20  & $-$1.40 &    1.30  &  1 \nl
BD +30 2611 &   PSK &  4275 & 0.80  & $-$1.40 &    2.10 & \nodata \nl
BD +52 1601 &   PSK &  4750 & 1.60  & $-$1.50 &    2.00  &  1 \nl
BD +54 1323 &   PSK &  5300 & 2.50  & $-$1.50 &    1.70 & \nodata \nl
BD +58 1218 &   PSK &  5000 & 2.20  & $-$3.00 &    1.00 & \nodata \nl
\tablenotetext{a}{Model parameters have been derived as follows:
PSK - Pilachowski et al. 1996; 
(PSK) - models have been derived following the methodology of PSK;
new - model parameters from PSK have been revised to obtain a better fit
to the Fe and Ti equivalent widths; McW - McWilliam et al. 1995b;
SKPL91 - Sneden et al. 1991; PSB - Pilachowski et al. 1993.}
\tablecomments{1.  Abundances derived from red Ba II $\lambda$6141 spectra only. \\
2.  Eu abundances derived from equivalent widths from Gilroy et al. (1988).}
\enddata
\end{deluxetable}